\documentclass[aps,pra,twocolumn,nofootinbib,english]{revtex4-1}
\usepackage[T1]{fontenc}
\usepackage[latin9]{inputenc}
\usepackage{babel}
\usepackage{mathtools}
\usepackage{amsmath}
\usepackage{amsthm}
\usepackage{amssymb}
\usepackage[unicode=true,pdfusetitle,
bookmarks=true,bookmarksnumbered=false,bookmarksopen=false,
breaklinks=false,pdfborder={0 0 1},backref=false,colorlinks=false]
{hyperref}
\usepackage{xcolor}

\makeatletter
\theoremstyle{plain}
\newtheorem*{thm*}{\protect\theoremname}
\theoremstyle{remark}
\newtheorem*{claim*}{\protect\claimname}

\usepackage{xfrac}
\usepackage{dsfont}
\usepackage{mathrsfs}
\newcommand{\id}{\mathds{1}}
\newcommand{\ve}{\text{vec}\, }
\newcommand{\veT}{\text{vec}^T }
\newcommand{\tr}{\text{Tr}}
\newcommand{\hess}{\mathscr{H}}
\newcommand{\eq}[1]{\begin{equation}\begin{aligned}#1\end{aligned}\end{equation}}
\newcommand{\had}{\hat{a}^\dagger\vphantom{n}}
\newcommand{\ha}{\hat{a}\vphantom{n}}
\newcommand{\iu}{\text{i}}
\newcommand{\eu}{\text{e}}
\newcommand{\ket}[1]{\left|#1\right\rangle}
\newcommand{\bra}[1]{\left\langle#1\right|}

\makeatother

\providecommand{\theoremname}{Theorem}
\providecommand{\claimname}{Claim}

\begin{document}
	
	\title{Perturbative expansion of entanglement negativity using patterned matrix calculus}

	\author{Jesse C. Cresswell}
	\email{jcresswe@physics.utoronto.ca}
	\author{Ilan Tzitrin}
	\email{itzitrin@physics.utoronto.ca}
	\author{Aaron Z. Goldberg}
	\email{goldberg@physics.utoronto.ca}
	
	\affiliation{Department of Physics, University of Toronto, Toronto, Ontario M5S 1A7, Canada}
	
	\begin{abstract}
		Negativity is an entanglement monotone frequently used to quantify entanglement in bipartite states. Because negativity is a non-analytic function of a density matrix, existing methods used in {the} physics literature are insufficient to compute its derivatives. To this end we develop techniques in the calculus of complex, patterned matrices and use them to conduct a perturbative analysis of negativity in terms of arbitrary variations of the density operator. The result is an easy-to-implement expansion that can be carried out to all orders. On the way we provide convenient representations of the partial transposition map appearing in the definition of negativity. Our methods are well-suited to study the growth and decay of entanglement in a wide range of physical systems, including the generic linear growth of entanglement in many-body systems, and have broad relevance to many functions of quantum states and observables.
	\end{abstract}
	
	\maketitle
	
	
	\section{Introduction}\label{sec:intro}

	In 1935, Einstein, Podolsky, and Rosen imagined a composite quantum
	state that did not admit a complete local description. 
	In this kind of state, outcomes of measurements performed on the subsystems were perfectly anti-correlated regardless of the chosen measurement basis. Schr\"{o}dinger
	shortly gave this remarkable feature of the quantum formalism the name
	\emph{entanglement}, and the notion continues to be the subject of
	extensive theoretical study and experiment \cite{Horodecki2009}, with applications like quantum teleportation \cite{Bennett1993,Bouwmeester1997,Barrett2004,Riebe2004}, quantum-enhanced metrology \cite{Mitchell2004, Pezze2009,giovannetti2011advances,gross2012nonlinear}, and quantum cryptography \cite{Ekert1991,Gisin2002,Curty2004,Tang2014}.
	The inquiry continues with our paper, {which gives} insight
	into how entanglement evolves as the state of a bipartite system is varied. 
	
	More concretely, for any Hilbert spaces $\mathcal{H}^{A}$ and $\mathcal{H}^{B}$ with dimensions $d_{A}$ and $d_{B}$, respectively,
	we can define the composite space $\mathcal{H}{}^{AB}\coloneqq\mathcal{H}^{A}\otimes\mathcal{H}^{B}$. Let $\rho$ be a density operator acting on $\mathcal{H}^{AB}$. If there exist density operators $\left\{ \rho_{i}^{A}\right\} $ and $\left\{ \rho_{i}^{B}\right\} $ acting on the Hilbert spaces $\mathcal{H}^A$ and $\mathcal{H}^B$ alone,
	and probabilities $\left\{ 0<p_{i}\leq1\right\} $ so that $\rho=\sum_{i}p_{i}\rho_{i}^{A}\otimes\rho_{i}^{B}$,
	we say that $\rho$ is \emph{separable}; if not, $\rho$ is \emph{entangled}.
	
	This definition seldom makes it apparent whether a given quantum state $\rho$ is separable, and moreover gives no criteria for comparing the degree to which different states are entangled. This motivated the development of a number of metrics to appropriately quantify and classify entanglement. The distillable entanglement \cite{Bennett1996}, entanglement of formation \cite{Wootters1998}, and various quantities based on entropies (e.g. the R\'enyi entropies \cite{Renyi1961}) satisfy certain properties that make them
	attractive \emph{entanglement measures} \cite{Plenio2005}. Among those properties are \emph{monotonicity},
	which ensures that entanglement does not increase under local
	operations and classical communication (LOCC); \emph{(sub)additivity}, which requires that the entanglement of a composite system is (less than or) equal to the sum of the entanglement in the subsystems; and \emph{convexity}, which means that they are convex functions of the density operator.
	
	In this paper we focus on
	the \emph{negativity, }a measure whose origins can roughly be traced
	to 1996, with Peres' Positive Partial Transpose (PPT) condition \cite{Peres1996}: if a
	bipartite state is separable, the transpose taken with respect to
	any subsystem is positive. The PPT condition, stronger and more efficient
	than entropic criteria based on the R\'enyi entropies $S_\alpha(\rho)\coloneqq\tfrac{1}{1-\alpha}\log \tr \left(\rho^\alpha\right)$ \cite{Vollbrecht2002},
	was shown by the Horodeckis to be sufficient for the separability
	of $2\otimes2$ and $2\otimes3$ systems \cite{Horodecki1996}. We
	summarize this as follows:
	\begin{thm*}[Peres-Horodecki Criterion]
		\emph{Let $T_{B}\left(\rho\right)\coloneqq\rho^{T_{B}}\coloneqq \left(\id\otimes T\right)\rho$
			be the }partial transposition map\emph{ with respect to system } $B$.
		\emph{Then $\rho\text{ is separable}\implies\rho^{T_{B}}\geq 0.$ Furthermore,
			if $(d_{A},d_{B})\in\{(2,2),(2,3)\}$,
			then $\rho^{T_{B}}\geq0\implies\rho\text{ is separable}$.
		}
		\label{thm:PHC}
	\end{thm*}
	From the Peres-Horodecki Criterion emerges the negativity, $\mathcal{N}$,
	of $\rho$\emph{ }\cite{Zyczkowski1998,Vidal2001}, which encodes
	the degree to which the partial transpose of $\rho$ is negative. Negativity is defined as
	\begin{equation}
	\mathcal{N}\left(\rho\right)\coloneqq\sum_{\lambda<0}\left|\lambda\left(\rho^{T_{B}}\right)\right|,
	\label{eq:negativity-eigvals}
	\end{equation}
	where the $\lambda$'s are the eigenvalues of $\rho^{T_{B}}$. Negativity is monotonic under LOCC  and convex but not additive \cite{Plenio2005a}. Although there exist entangled states for which $\mathcal{N}$ vanishes, negativity has an important advantage over other measures\footnote{By most definitions, entanglement measures ought to reduce to the \emph{entanglement entropy} \cite{Bennett1996a, Plenio2005} in the pure state case; negativity does not. Nevertheless, we alternate between calling it a monotone and measure depending on the subsystem dimensions.} in that it is easily computable, even for mixed states \cite{Huang2014}. For some
	purposes it is useful to define the \emph{logarithmic negativity},
	$E_{\mathcal{N}}\left(\rho\right)\coloneqq\log_{2}\left[2\mathcal{N}\left(\rho\right)+1\right]$, which is monotonic and additive but not convex  \cite{Plenio2005a}. Relationships between the logarithmic negativity, distillable entanglement, entanglement cost, and teleportation capacity have been demonstrated in the literature \cite{Vidal2001, Audenaert2003}.
	
	The dynamics of entanglement, studied initially in quantum-optical systems, has become an active area of research
	in many-body systems \cite{Audenaertetal2002,Anders2008,EislerZimboras2014}, condensed-matter physics \cite{HelmesWessel2014,EislerZimboras2015,Shermanetal2016,Shimetal2018}, and quantum field
	theories \cite{Calabreseetal2012,Coser2014,Chaturvedietal2018}. Perturbative approaches to entanglement dynamics have revealed a universal timescale characterizing the growth of entanglement
	in initially pure, separable states under unitary evolution as measured by the purity of the reduced density matrix \cite{Kim1996, Yang2017}, and by the R\'enyi entropies \cite{Cresswell2017a}.
	
	Although negativity has been studied for two decades, there is, to our knowledge, no general perturbative expansion available. This omission is likely owed to three nuances in the differentiation of negativity: the matrix representation of the partial transposition map, issues of non-differentiability of the trace norm, and the computation of the trace norm's derivative, which requires a careful consideration of the calculus of complex matrices with patterns. 
	
	To expand on these difficulties, we recast negativity in terms of the \emph{trace norm}\footnote{The trace norm, also known as the \textit{nuclear norm}, is one of the \textit{Schatten norms} and one of the \textit{Ky Fan norms}.} $\left\Vert X\right\Vert _{1}\coloneqq\text{Tr}\sqrt{X^{\dagger}X}$ of a matrix, $X$, so that
	\begin{equation}
	\mathcal{N}\left(\rho\right)\coloneqq\frac{1}{2}\left(\left\Vert \rho^{T_{B}}\right\Vert _{1}-1\right).
	\end{equation}
	A perturbative expansion of $\mathcal{N}\left(\rho\right)$ as $\rho(\mu)$ varies with respect to some parameter $\mu$ requires a derivative that we schematically write as 
	\begin{equation}\label{eq:SchematicDerivative}
	\frac{d\mathcal{N}}{d \mu}=\frac{1}{2}\frac{\partial\left\Vert \rho^{T_{B}}\right\Vert _{1}}{\partial\rho^{T_{B}}}\frac{\partial\rho^{T_{B}}}{\partial\rho}\frac{\partial\rho}{\partial \mu}.
	\end{equation}
	In the following sections we will be more precise about how these derivatives are defined and multiplied. The last factor, ${\partial\rho}/{\partial \mu}$, will depend on the application at hand, but for concreteness we will mainly consider time evolution, with ${\partial\rho}/{\partial t}$ governed by the von Neumann dynamics of a closed system or the Lindblad dynamics of an open system. It should be understood that more general variations can be treated in the same way. The middle factor requires an explicit expression for the action of the partial transposition map on the density matrix. Some forms are available in the literature, but in what follows we work in the vectorized representation where $\rho$ and $\rho^{T_B}$ are vectors in $\mathbb{C}^{(d_Ad_B)^2}$. The linear map $T_{B}=\id\otimes T$ has a simple representation in this vector space that is easy to implement numerically, while also providing a form for ${\partial\rho^{T_{B}}}/{\partial\rho}$. 
	
	The more challenging factor to understand is the derivative of the trace norm with respect to its argument. One complication is that the trace norm is only differentiable if its argument is invertible \cite{Watson1992}. At singular arguments one has the notion of a \emph{subdifferential}, to which we return in the discussion. Otherwise, we assume invertibility of $\rho^{T_B}$ throughout. Furthermore, the derivative ${\partial\left\Vert \rho^{T_{B}}\right\Vert _{1}}/{\partial\rho^{T_{B}}}$ is taken with respect to a Hermitian matrix whose elements are not independent variables. This is perilous; it is necessary to represent $\rho^{T_B}$ in terms of a set of independent variables before differentiating, which is the main notion behind patterned matrix calculus \cite{Hjorungnes2011}. Generally, we refer to any matrix whose elements are not independent variables as a \emph{patterned} matrix, some other examples of which are symmetric, unitary, or diagonal matrices. Patterned matrix calculus is an underexplored branch of mathematics that we find to be crucial in the perturbative analysis of negativity.

	In this paper, we address the nuances just mentioned to further our understanding of entanglement
	dynamics by way of negativity.
	In Section \ref{sec:Dynamics},
	we provide a means by which to compute the perturbative expansion of negativity.
	To this end, we offer new matrix representations of the partial transposition
	map, along with explicit computations of the first and second derivatives
	of the trace norm with respect to complex, patterned arguments. The techniques we develop can be straightforwardly carried out to any order in the expansion for negativity.
	In Section \ref{sec:systems}, we apply our results to several physical
	systems with illustrative differences and compare them with the behaviour of other entanglement
	measures. In Section \ref{sec:Discussion}, we discuss in greater
	detail the challenges of our approach, the validity of our assumptions, and the benefits and
	limitations of our results.

	\section{Perturbative expansion of negativity \label{sec:Dynamics}}
	Let us suppose that the density matrix, $\rho(t)$, of a quantum state and some number of its derivatives are known at a given time, $t_0$. To understand how the entanglement between two subsystems changes near $t_0$ due to the evolution of $\rho(t)$ we can expand the negativity as
	\begin{equation}\label{eq:expansion}
	\mathcal{N}(t)=\mathcal{N}(t_0)+\left.\frac{d \mathcal{N}}{d t}\right|_{t_0} (t-t_0)+\frac{1}{2}\left.\frac{d^2 \mathcal{N}}{dt^2}\right|_{t_0} (t-t_0)^2+\cdots .
	\end{equation}
	Our goal is to provide general expressions for the derivatives of negativity in this expansion.\footnote{The logarithmic negativity, $E_{\mathcal{N}}\left(\rho\right)\coloneqq\log_{2}\Vert\rho^{T_B}\Vert_{1}$, can be treated analogously.} In this section we offer explicit expressions for the first and second derivatives using a method that can be carried out systematically to any desired order of precision.
	
	\subsection{Vectorization formalism and the partial transposition map}
	\label{subsec:ptranspose}
	
	Expressions like \eqref{eq:SchematicDerivative} are cumbersome to work with since $\partial \rho^{T_B}/\partial \rho$ represents a four-dimensional array that must be contracted against two matrices to produce a scalar. To avoid such complications we prefer to work in the vectorization formalism, where we represent the state $\rho$ as a column vector, $\ve\rho$. In general, the vectorization operation, $\text{vec}$, stacks the columns of an $m\times n$ matrix into an $mn\times 1$ vector. It admits two useful identities that we rely on, namely
	\begin{equation}
	\ve\left(ABC\right) =\left(C^{T}\otimes A\right)\ve B\label{eq:vecABC},
	\end{equation}
	where $A,B,C$ are any three compatible matrices, and
	\begin{equation}
	\tr\left(A^T B\right) =(\ve ^T A) \ve B,\label{eq:TrAB}
	\end{equation}
	where $A,B$ are the same size and $\ve ^T A \coloneqq (\ve A)^T$ \cite{Magnus1985, Magnus2007}. In this formalism we can rewrite the first derivative of negativity as
	\begin{equation}\label{eq:FirstDerivative}
	\frac{d\mathcal{N}}{d t}=\frac{1}{2}\frac{\partial\left\Vert \rho^{T_{B}}\right\Vert _{1}}{\partial\,\veT\rho^{T_{B}}}\frac{\partial\,\ve\rho^{T_{B}}}{\partial\,\veT\rho}\frac{\partial\,\ve\rho}{\partial t}.
	\end{equation}
	With this simple change in notation the middle factor is an ordinary matrix multiplied by two vectors. To demonstrate the convenience of this formalism we first consider how the standard matrix transposition map can be represented, before applying it to the partial transposition and determining $\partial\,\ve\rho^{T_{B}}/{\partial\,\veT\rho}$.

	Let $\phi$ be a superoperator, i.e., a linear map from the space of $m\times n$ matrices to the space of $q\times r$ matrices. Due to linearity, $\phi$ can always be written as
	\begin{equation}
	\phi(X)=\sum_i A_i X B_i,
	\end{equation}
	where $X$, the $A_i$ and the $B_i$ are $m \times n$, $q\times m$, and $n\times r$ matrices, respectively. If $X$ is instead vectorized as $\ve X$, then there is a related operator, $M_\phi$, acting on the space of $mn\times 1$ vectors, and represented by the $qr\times mn$ matrix
	\begin{equation}\label{eq153}
	M_\phi=\sum_i  B_i^T\otimes A_i.
	\end{equation}
	There is no restriction on the number of terms in each sum, so these representations will not be unique. The $M_\phi$ operator can be viewed as implementing the linear map $\phi$ on the vectorized space of matrices,
	\begin{equation}\label{eq:vectorizedsupop}
	\ve \phi(X)= M_\phi \ve X,
	\end{equation}
	which follows from identity \eqref{eq:vecABC}. This demonstrates the existence of an isomorphism\footnote{This is distinct from the Choi-Jamiolkowski isomorphism.} between the space of superoperators acting on $\mathds{C}^{m,n}$ and those acting on $\mathds{C}^{mn}$ \cite{Havel2002, Zyczkowski2004}.
	
	Matrix transposition is a linear operation, meaning we can define a \emph{commutation matrix}, $K_{mn}$, such that
	\begin{equation}
	\ve X^{T}=K_{mn}\ve X,\label{eq:kmat}
	\end{equation}
	where $K_{mn}$ has dimensions $mn \times mn$. $K_{mn}$ is a symmetric permutation matrix satisfying the useful identities
	\begin{align} 
	(K_{mn})^{2} &= \id,\\
	\label{Kcommid}
	K_{mq}\left(X\otimes Y\right) &= \left(Y\otimes X\right)K_{nr},
	\end{align}
	where $Y$ is $q\times r$. In the following we only deal with square matrices and hence use $K \coloneqq K_{nn}$ for ease of notation. The commutation matrix has a simple representation of the form \eqref{eq153} in the standard basis of matrices $\left\{ J^{ij}_{n}\right\}$, where $J^{ij}_n$ is the $n\times n$ \emph{single-entry matrix} defined through
	$\left(J^{ij}_n\right)_{kl}=\delta_{ik}\delta_{jl}$. In terms of these elements
	\begin{equation}\label{eq:commutation}
	K=\sum_{i,j=1}^n J^{ji}_n\otimes J^{ij}_n.
	\end{equation}

	In the same vein as \eqref{eq:kmat}, we can represent the partial transposition map as a linear superoperator acting on the space of $\left(d_{A}d_{B}\right)^{2}\times 1$ column vectors,
	\begin{equation}
	\ve \rho^{T_{B}}=K_B\ve \rho ,
	\end{equation}
	where the subscript indicates the subsystem to be transposed. We call $K_{B}$ a \emph{partial commutation matrix}. It is $\left(d_{A}d_{B}\right)^{2}\times \left(d_{A}d_{B}\right)^{2}$ and is self-inverse. The advantage of this formalism is that we can immediately identify the middle factor in \eqref{eq:FirstDerivative} as
	\begin{equation}
	\frac{\partial\,\ve \rho^{T_{B}}}{\partial\,\veT \rho}=K_B,\label{eq:drhoTBdrho}
	\end{equation}
	which follows when we observe that the partial commutation matrix is constant. To establish this we can investigate the form of $K_B$, again working with standard basis elements. We find
	\begin{equation}
	K_B=\sum_{i,j=1}^{d_{B}}\left(\id_{d_{A}}\otimes J_{d_{B}}^{ji}\right)\otimes\left(\id_{d_{A}}\otimes J_{d_{B}}^{ij}\right).\label{eq:supop}
	\end{equation}
	We note that $K_B$ is a constant matrix that depends only on the dimensions $d_A$, and $d_B$. Furthermore, when $d_A=1$, then $K_B=K$, as expected. In App. \ref{sec:superoperator} we show how \eqref{eq:supop} can be obtained through the action of the partial transposition map on the standard basis, and we also present a convenient form for $K_B$ by identifying its eigenvectors.  For other representations of the transposition and partial transposition maps, see, for example, \cite{Jaroslaw2011}.
	
	With this brief introduction to aspects of algebra on vectorized matrices, we now turn to calculus in order to identify ${\partial\left\Vert \rho^{T_{B}}\right\Vert _{1}}/{\partial\,\veT\rho^{T_{B}}}$, the remaining factor in \eqref{eq:FirstDerivative}.

	\subsection{First derivative of the trace norm \label{subsec:trnormder}}
	
	In this section we use $X$ to represent an unpatterned $n\times n$ matrix of complex variables. The unpatterned derivatives of a scalar function $g(X,X^*)$,
	\begin{equation}
	D_X g\coloneqq \frac{\partial g}{\partial\,\veT X},\quad \text{and} \quad D_{X^*} g\coloneqq \frac{\partial g}{\partial\,\veT X^*},
	\end{equation}
	are found by expressing the differential of $g$ in the form
	\begin{equation}\label{eq:differentialTrNorm}
	d g =(D_X g) d\ve X+(D_{X^*} g) d\ve X^*,
	\end{equation}
	and reading off the prefactors of $d\ve X$ and $ d\ve X^*$. The differentials $d\ve X$ and $d\ve X^*$ are taken to be independent, and when $g$ is differentiable the unpatterned derivatives are unique \cite{Magnus1985,Hjorungnes2007}.

	As discussed in Sec.\ref{sec:intro}, $\rho^{T_B}$ is Hermitian, meaning its matrix elements are interdependent and special care must be taken in defining a derivative with respect to it. With some effort, the derivative of a scalar function with respect to a Hermitian argument $A$ can be found in terms of the unpatterned derivatives as in \cite{Hjorungnes2011}:
	\begin{equation}\label{eq:DTrNorm}
	D_{A} g\coloneqq\frac{\partial g}{\partial\,\veT A}=(D_X g)|_{X=A}+ (D_{X^*} g)|_{X=A} K,
	\end{equation}
	where $(D_X g)|_{X=A}$ means the Hermitian pattern is applied after the unpatterned derivative has been computed, and $K$ is the commutation matrix from \eqref{eq:commutation}. App. \ref{sec:Acomp_pat} contains an abstract summary of our general approach to patterned derivatives {which can be used to establish \eqref{eq:DTrNorm} rigourously. For a briefer derivation, we note that the differentials of $A$ are not independent, since $d\ve A^{*}=d\ve A^{T}=Kd\ve A$. Hence, when we apply the Hermitian pattern, \eqref{eq:differentialTrNorm} becomes
		\begin{equation}
		dg=
		\left[D_{X}g+\left(D_{X^{*}}g\right)K\right]_{X=A}d\ve A,
		\end{equation}
		and we identify the patterned derivative in \eqref{eq:DTrNorm}. We note that $D_{A^*}g$ can be found similarly, but is not independent from $D_{A}g$ since $D_{A^*}g=(D_{A}g) K$.}

	The trace norm $\Vert \rho^{T_B}\Vert_{1}$ is a scalar function of a Hermitian matrix, so in order to find its derivative we start by computing the differential of $\Vert X\Vert_{1}$ as in \eqref{eq:differentialTrNorm}. We take $X$ to be invertible so that $\Vert X\Vert_{1}$ is differentiable, and define $\left|X\right|\coloneqq\sqrt{X^{\dagger}X}$, so that
	$\left\Vert X\right\Vert _{1}=\text{Tr}\left|X\right|$. Then, by definition,
	$\left|X\right|\left|X\right|=X^{\dagger}X$, and we can take the
	differential of both sides to obtain
	\begin{equation}
	\left(d\left|X\right|\right)\left|X\right|+\left|X\right|d\left|X\right|=X^{\dagger}dX+\left(dX^{\dagger}\right)X\label{eq:firstdif}.
	\end{equation}
	Multiplying by $\left|X\right|^{-1}$ on the left and taking the trace allows us to isolate $\tr \left(d\left\vert X\right\vert\right)$
	\begin{equation}
	2\tr\left(d\left|X\right|\right)=\tr\left(\left\vert X\right\vert^{-1} X^\dag dX \right)+\tr\left(X \left\vert X\right\vert^{-1} dX^\dag\right).
	\end{equation}
	The differential operator $d$ commutes with both the trace and $\text{vec}$ operations, so this equation actually gives the differential of the trace norm, $\tr\left(d\left\vert X\right\vert\right)=d \Vert X\Vert_{1}$. Identity \eqref{eq:TrAB} allows us to express $d\Vert X\Vert_{1}$ in the required form \eqref{eq:differentialTrNorm},
	\begin{equation}
	\begin{aligned}
	d \Vert X\Vert_{1}=&\frac{1}{2}\veT\left[X^{*}\left(\left|X\right|^{-1}\right)^{T}\right]d\ve X\\
	&+\frac{1}{2}\veT\left[\left(\left|X\right|^{-1}\right)^{T}X^{T}\right]Kd\ve X^{*}\label{eq:firstdifferential},
	\end{aligned}
	\end{equation}
	where we have also used \eqref{eq:kmat} for $d\ve X^\dag$. Now we can identify the derivatives with respect to $X$ and $X^*$ as
	\begin{align}
	    \begin{aligned}
	    D_X \Vert X\Vert_{1}&=\frac{1}{2}\veT\left[X^{*}\left(\left|X\right|^{-1}\right)^{T}\right], \\ D_{X^*} \Vert X\Vert_{1}&=\frac{1}{2}\veT\left[\left(\left|X\right|^{-1}\right)^{T}X^{T}\right]K.\label{eq:unpatder}
	    \end{aligned}
	    \end{align}
	The derivative with respect to a Hermitian argument follows by substituting \eqref{eq:unpatder} in \eqref{eq:DTrNorm}, and recalling $K^2=\id$:
	\begin{align}
	\begin{aligned}
	D_A \Vert A\Vert_{1}&=\frac{1}{2}\veT\left[X^{*}\left(\left|X\right|^{-1}\right)^{T}+\left(\left|X\right|^{-1}\right)^{T}X^{T}\right]_{X=A}\\
	&=\veT\left[A^T\left(\left|A\right|^{-1}\right)^T\right]=\veT\left[A\left\vert A\right\vert^{-1}\right]K\label{eq:drhopatterned},
	\end{aligned}
	\end{align}
	where we note that $A$ commutes with $\left\vert A\right\vert^{-1}$. The matrix $A\left\vert A\right\vert^{-1}$ appearing in the derivative is the matrix extension of the sign function, defined such that $A=\text{sign}(A)|A|$ \cite{Higham2008}.

	The derivative simplifies even further when we use the eigendecomposition for $\rho^{T_B}=U\Lambda U^\dag$, where $\Lambda$ contains the eigenvalues of $\rho^{T_B}$ in decreasing order, 
	\begin{equation}
	\frac{\partial\left\Vert \rho^{T_{B}}\right\Vert _{1}}{\partial\,\veT\rho^{T_{B}}}=D_{\rho^{T_B}}\left\Vert \rho^{T_{B}}\right\Vert _{1}=\veT\left( U^{*}\text{sign}\left(\Lambda\right)U^{T}\right).
	\label{eq:patder}
	\end{equation}
	
	We would like to emphasize that the derivative taken with respect to a Hermitian argument \eqref{eq:patder} is twice the unpatterned derivative \eqref{eq:unpatder}, which shows the indispensability of patterned matrix calculus for understanding derivatives of negativity. Our results \eqref{eq:supop} and \eqref{eq:patder} can be combined to give the first derivative of negativity \eqref{eq:FirstDerivative}. Next, we compute the second derivative explicitly, and then proceed to show how the perturbative expansion can be carried out to any order.

	\subsection{Second derivative of the trace norm \label{subsec:trnormhessian}}
	
	Taking another derivative of \eqref{eq:FirstDerivative} and noting that $K_B$ is constant gives
	\begin{align}\begin{aligned}\label{eq:secondDerivative}
	\frac{d^2 \mathcal{N}}{dt^2}=&\frac{1}{2}\left(\frac{\partial\,\ve \rho^{T_B}}{\partial\,\veT \rho}\frac{\partial\,\ve\rho}{\partial t}\right)^T\frac{\partial}{\partial\,\veT \rho^{T_B}}\left(\frac{\partial\left\Vert\rho^{T_B}\right\Vert_{1}}{\partial\,\veT \rho^{T_B}}\right)^T
	\\&\times
	\left(\frac{\partial\,\ve \rho^{T_B}}{\partial\,\veT \rho}\frac{\partial\,\ve\rho}{\partial t}\right)\\
	&+\frac{1}{2}\frac{\partial\left\Vert\rho^{T_B}\right\Vert_{1}}{\partial\,\veT \rho^{T_B}}\frac{\partial\,\ve \rho^{T_B}}{\partial\,\veT \rho}\frac{\partial^2\,\ve\rho}{\partial t^2}.
	\end{aligned}\end{align}
	The second line can be computed using the above results if $\partial^2 \rho/\partial t^2$ is known, but the first line involves the Hessian of $\Vert\rho^{T_B}\Vert_{1}$ with respect to a Hermitian argument,
	
	\begin{equation}\label{eq:HermHessian}
	\hess_{\rho^{T_B},\rho^{T_B}}\left(\Vert\rho^{T_B}\Vert_{1}\right)\coloneqq \frac{\partial}{\partial\,\veT \rho^{T_B}}\left(\frac{\partial\left\Vert\rho^{T_B}\right\Vert_{1}}{\partial\,\veT \rho^{T_B}}\right)^T.
	\end{equation}
	To our knowledge, Hessians with respect to patterned matrices have not been discussed in the literature. {Therefore, in the following we present a detailed discussion of such Hessians in general before applying our new methods to the trace norm to compute \eqref{eq:HermHessian}. 
	}
	
	As with Jacobians \eqref{eq:differentialTrNorm}, unpatterned Hessians are defined through the differential,\footnote{We follow the conventions in \cite{Hjorungnes2011}.}
	\eq{\label{eq:seconddifferentialTrNorm}
	d^2 g = & d\veT X \left[ \hess_{X,X}(g) \right] d\ve X+d\veT X^* \\&\times
	\left[ \hess_{X,X^*}(g) \right] d\ve X
	+d\veT X \left[ \hess_{X^*,X}(g) \right] d\ve X^*\\&
	+d\veT X^*  \left[ \hess_{X^*,X^*}(g) \right] d\ve X^*,}
	where we take $\hess_{X^*,X}(g)\coloneqq D_{X^*}(D_X g)^T\coloneqq\frac{\partial}{\partial\,\veT X^*}\left[\frac{\partial g}{\partial\,\veT X}\right]^T$, etc., and note that second differentials of the variables $X$ and $X^{*}$ are zero by definition. Since the four Hessians here are not independent, there is some freedom to exchange terms in the form \eqref{eq:seconddifferentialTrNorm}. To compute the Hessians, one must write the second differential of $g(X,X^*)$ as
	\begin{align}\begin{aligned}\label{eq:Aexpansion}
	d^2 g {=} & d\veT X (B_{10})\ d\ve X{+}d\veT X^* (B_{00})\ d\ve X\\
	&{+}d\veT X (B_{11})\ d\ve X^*{+}d\veT X^* (B_{01})\ d\ve X^*.
	\end{aligned}\end{align} 
	Since partial derivatives commute, the $\hess_{X,X}$ and $\hess_{X^*,X^*}$ Hessians should be symmetric, while $\hess_{X,X^*}$ and $\hess_{X^*,X}$ should be transposes of one another. One can conclude that
	\begin{align}
	\hess_{X,X}(g)&=\frac{1}{2}\left(B_{10}+B_{10}^T\right),\label{eq:unpatHess1}\\
	\hess_{X^*,X^*}(g)&=\frac{1}{2}\left(B_{01}+B_{01}^T\right),\label{eq:unpatHess2}\\
	\hess_{X,X^*}(g)&=\frac{1}{2}\left(B_{00}+B_{11}^T\right)=[\hess_{X^*,X}(g)]^T.\label{eq:unpatHess3}
	\end{align}


	
	When the argument of $g(A,A^*)$ is a Hermitian matrix, this procedure must be modified to account for the interdependence of $A$ and $A^\dag$. We can define a Hessian with respect to a Hermitian argument by noting
	that the patterned Jacobian for scalar $g$, Eq. \eqref{eq:DTrNorm}, also applies to a vector function $\boldsymbol{g}$,
	since it applies elementwise:
	\begin{equation}
	D_{A}\boldsymbol{g}=\left[D_{X}\boldsymbol{g}+\left(D_{X^{*}}\boldsymbol{g}\right)K\right]_{X=A}.\label{eq:hermvecder}
	\end{equation}
	Then, the Hessian of $g\left(A,A^{*}\right)$ with respect to Hermitian $A$ can be found by applying \eqref {eq:hermvecder} to the Jacobian $\boldsymbol{g}=\left(D_{A}g\right)^{T}$, resulting in
	\eq{\label{eq:hessscal}
	\hess_{A,A}(g)  :=&D_{A}\boldsymbol{g} =\left[D_{X}\left(D_{X}g\right)^{T}+K D_{X}\left(D_{X^{*}}g\right)^{T}\right.\\&\left.
	+D_{X^{*}}\left(D_{X}g\right)^{T}K+KD_{X^{*}}\left(D_{X^{*}}g\right)^{T}K\right]_{X=A} 
	\\
	 =&\hess_{X,X}(g)\vert_{X=A}+K\hess_{X,X^{*}}(g)\vert_{X=A}\\&
	+\hess_{X^{*}X}(g)\vert_{X=A}K+K\hess_{X^{*}X^{*}}(g)\vert_{X=A}K.
	}
	
	One way to gauge the correctness of this result is to use expression \eqref{eq:seconddifferentialTrNorm}
	for the second differential of a scalar function $g(X,X^*)$. Then, letting $X\to A$ and recalling $d\ve A^{*}=Kd\ve A$ gives
	\begin{align}
	\begin{aligned}
	d^{2}g  =&d\veT A[\hess_{X,X}(g)\vert_{X=A}+K\hess_{X,X^{*}}(g)\vert_{X=A}\\
	&{+}\hess_{X^{*}X}(g)\vert_{X=A}K+K\hess_{X^{*}X^{*}}(g)\vert_{X=A}K]d\ve A. \label{eq:hessalt}
	\end{aligned}
	\end{align}
	We can see that the part of expression \eqref{eq:hessalt} in square brackets \textendash{}
	defined to be $\hess_{A,A}(g)$ \textendash{} matches Eq. \eqref{eq:hessscal}. This result can also be confirmed using the formal calculus described in App. \ref{sec:Acomp_pat}. Whereas there are three independent Hessians of $g$ with respect to combinations of $\{X,X^*\}$, there is only one independent Hessian in the Hermitian case. The Hessians are related by $\hess_{A,A^*}(g)=K\hess_{A,A}(g)$, $\hess_{A^*,A}(g)=\hess_{A,A}(g)K$, and $\hess_{A^*,A^*}(g)=K\hess_{A,A}(g)K$.

	We now use our result \eqref{eq:hessscal} to compute \eqref{eq:HermHessian} from the unpatterned Hessians of the trace norm. Taking the differential of both sides of Eq. \eqref{eq:firstdif} and noting $d^{2}X=0=d^{2}X^{\dagger}$ gives us
	\begin{equation}
	\left(d^{2}\left|X\right|\right)\left|X\right|+\left|X\right|d^{2}\left|X\right|=2dX^{\dagger}dX-2\left(d\left|X\right|\right)^2.
	\label{eq:doublediff}
	\end{equation}
	Once again we left-multiply by $|X|^{-1}$ and take the trace,
	\begin{equation}
	{\tr}( d^{2}\left|X\right|){=}\tr[(dX^{\dagger}){\left(dX\right)}{\left\vert X\right\vert^{-1}]}-\tr[\left(d\left\vert X\right\vert\right)\left|X\right|^{-1}d\left\vert X\right\vert],
	\end{equation}
	which becomes
	\eq{d^{2}\left\Vert X\right\Vert_{1}=&d\veT\left(X^{*}\right)\left[\left(\left\vert X\right\vert^{-1}\right)^{T}\otimes\id\right]d\ve X\\&
	-d\veT\left(\left|X\right|^T\right)\left[\id\otimes\left|X\right|^{-1}\right]d\ve \left|X\right|.\label{eq:d2tr}}
	To find $d\ve\left|X\right|$, we may vectorize both sides of Eq. \eqref{eq:firstdif} and use identity \eqref{eq:vecABC} which results in
	\begin{align}
	 \begin{aligned}
	   (\left|X\right|^{T}&\otimes\id+\id\otimes\left|X\right|) d\ve \left|X\right|\\
	=&\left(\id\otimes X^{\dagger}\right)d\ve X+\left(X^{T}\otimes\id\right)d\ve X^{\dagger}.
	  \end{aligned}
	\end{align}
	For compact notation, let us introduce the \emph{Kronecker sum} $A\oplus B \coloneqq A\otimes\id+\id\otimes B$ and define $X_{\oplus}\coloneqq X^{T}\otimes\id+\id\otimes X=X^{T}\oplus X$. Then, since $|X|_{\oplus}$ is invertible,\footnote{Notice
		that $
		\text{det}|X|_{\oplus}\geq\text{det}\left(\left|X\right|^{T}\otimes\id\right)+\text{det}\left(\id\otimes\left|X\right|\right)=2\left(\text{det}\left|X\right|\right)^{n}>0$,
		where we have used the fact that $\left|X\right|^{T}\otimes\id$ and
		$\id\otimes\left|X\right|$ are positive semidefinite and $\left|X\right|$
		is nonsingular. Hence $|X|_{\oplus}$ is invertible. Eq. \eqref{eq:firstdif} is a \emph{Sylvester equation} for which solvability conditions are known and met in our case. 
	}
	\eq{d\ve \left|X\right|=&\left(|X|_{\oplus}\right)^{-1}\left(\id\otimes X^{\dagger}\right)d\ve X\\&
	+\left(|X|_{\oplus}\right)^{-1}\left(X^{T}\otimes\id\right)Kd\ve X^{*}.\label{eq:dvec|X|}}
	Inserting this in Eq. \eqref{eq:d2tr} brings us to the desired form \eqref{eq:Aexpansion} from which we can read off the $B$ matrices, and combine them to form the Hessians in Eqs. \eqref{eq:unpatHess1} to \eqref{eq:unpatHess3}. 
	
	All that remains is to merge the unpatterned Hessians as in \eqref{eq:hessscal} to obtain the Hessian with respect to a Hermitian variable. 
	We show these computations in more detail in App. \ref{sec:simphess}. The result is
	\eq{\label{eq:patternedHessian}
	{\hess_{\rho^{T_{B}},\rho^{T_{B}}}}{\left(\left\Vert \rho^{T_{B}}\right\Vert _{1}\right)}=&\frac{1}{2}K\left[{\left(\left|\rho^{T_{B}}\right|^{-1}\right)_{\oplus}}{-}\rho_{\oplus}^{T_{B}}\left(\left\vert\rho^{T_{B}}\right\vert_{\oplus}\right)^{-1}\right.\\&\left.
	\times\left(\left\vert\rho^{T_{B}}\right\vert^{-1}\right)_{\oplus}\left(\left\vert\rho^{T_{B}}\right\vert_{\oplus}\right)^{-1}\rho_{\oplus}^{T_{B}}\right].
	}
	For computational efficiency we can simplify this expression in terms of the  eigendecomposition of $\rho^{T_{B}}$, as we did for the first derivative. We find
	\eq{\hess_{\rho^{T_{B}},\rho^{T_{B}}}\left(\left\Vert \rho^{T_{B}}\right\Vert _{1}\right) 
	&=K(U^*\otimes U)\left[
	\id-\text{sign}\Lambda\right.\\&\left.
	\otimes\ \text{sign}\Lambda\right]\left(\left|\Lambda\right|_{\oplus}\right)^{-1}(U^T\otimes U^\dag).\label{eq:spectHessian2}}
	This form provides additional insight into the behaviour of negativity since the Hessian vanishes when the eigenvalues of $\rho^{T_B}$ are all positive. This Hessian, along with the results of Secs. \ref{subsec:ptranspose} and \ref{subsec:trnormder}, allows the second derivative of negativity \eqref{eq:secondDerivative} to be written in terms of the density matrix and derived quantities.

	\subsection{Summary of method \label{subsec:summary}}
	
	In this section we summarize our results in an algorithm for computing the perturbative expansion for negativity \eqref{eq:expansion} to second order:
	
	\renewcommand{\theenumi}{\Roman{enumi}}

	(1) Determine the derivatives of the density matrix at $t=t_0$, e.g. from an equation of motion.
		
		(2) Construct the commutation matrix $K$ from \eqref{eq:commutation} and partial commutation matrix $K_B$ from \eqref{eq:supop} appropriate for subsystem dimensions $d_A,d_B$ from the basis of single-entry matrices $J_n^{ij}$.
		
		(3) Compute the eigendecomposition $\rho^{T_B}(t_0)=U \Lambda U^\dag$.
		
		(4) Use the above in the first derivative of negativity, found from \eqref{eq:FirstDerivative} using \eqref{eq:drhoTBdrho} and \eqref{eq:patder}:
		\begin{align}
		\left.\frac{d\mathcal{N}}{d t}\right|_{t=t_0}&=\frac{1}{2}\veT\left[U^{*}\text{sign}\left(\Lambda\right)U^{T}\right]K_B\ve\dot{\rho}\left(t_0\right).
		\label{eq:first derivative}
		\end{align}
		
		(5) Use the above in the second derivative of negativity, found from \eqref{eq:secondDerivative} using \eqref{eq:drhoTBdrho}, \eqref{eq:patder}, and additionally \eqref{eq:spectHessian2} for the patterned Hessian of the trace norm $\hess_{\rho^{T_{B}},\rho^{T_{B}}}\left(\left\Vert \rho^{T_{B}}\right\Vert _{1}\right)$. We summarize it as
		\begin{align}
		    \begin{aligned}
		    \label{eq:second derivative}
		{\left.\frac{d^{2}\mathcal{N}}{d t^{2}}\right|_{t=t_0}}{=} & \frac{1}{2}{\left[K_B\ve\dot{\rho}\left(t_0\right)\right]^{T}}{\hess_{\rho^{T_{B}},\rho^{T_{B}}}}{\left(\left\Vert \rho^{T_{B}}\right\Vert _{1}\right)}{\left[K_B\ve\dot{\rho}\left(t_0\right)\right]}\\
		& +\frac{1}{2}\veT\left[U^{*}\text{sign}\left(\Lambda\right)U^{T}\right]K_B\ve\ddot{\rho}\left(t_0\right).
		    \end{aligned}
		\end{align}

	In light of Sections \ref{subsec:trnormder} and \ref{subsec:trnormhessian}, it is clear that higher differentials $d^{n} \Vert X\Vert_{1}$, with $n\geq 3$, can be computed iteratively by solving the equation $ d^n(|X| |X|)=d^n(X^\dag X)=0$ for the differentials $d^n\ve |X|$. Each such equation takes the form
	\begin{equation}
	(d^n\left\vert X\right\vert) \left\vert X\right\vert+\left\vert X\right\vert d^n\left\vert X\right\vert=C_n,\end{equation}
	where $C_n$ only contains differentials of order less than $n$, and each equation can be solved as in \eqref{eq:dvec|X|},
	\begin{equation}
	d^n\ve \left\vert X\right\vert=\left(\left\vert X\right\vert_{\oplus}\right)^{-1}\ve C_n.
	\end{equation}
	In terms of lower-order differentials of $\left\vert X\right\vert$, then, 
	\begin{equation}\label{eq:ndifferential}
	d^n\left\Vert X\right\Vert_{1}=\frac{1}{2} \tr (\left\vert X\right\vert^{-1} C_n).
	\end{equation}
	Finally, the form of higher-order derivatives of $\left\Vert \rho^{T_B}\right\Vert_{1}$ with respect to its Hermitian argument can be generalized from the methods in App. \ref{sec:Acomp_pat} and read off from \eqref{eq:ndifferential}.
	This extends the steps in Section \ref{subsec:trnormhessian} for a perturbative expansion of negativity to any order. 
	
	We conclude this section by noting that, for certain classes of systems,  all terms involving higher derivatives of the trace norm vanish. In these cases we have
	\begin{equation}
	\frac{d^{n}\mathcal{N}}{d t^{n}}=\frac{1}{2}\veT\left[U^{*}\text{sign}\left(\Lambda\right)U^{T}\right]K_B\ve\left.\frac{\partial^{n}\rho}{\partial t^{n}}\right\vert_{t=t_0},
	\end{equation}
	and the expansion \eqref{eq:expansion} resums to
	\eq{\label{eq:specialexpansion}
	{\mathcal{N}\left(t\right)}{=}&{\mathcal{N}\left(t_{0}\right)}{+}{\frac{1}{2}}{\left.\left(\text{vec}^{T}\left[U^{*}\text{sign}\left(\Lambda\right)U^{T}\right]K_B\right)\right\vert_{t=t_{0}}}{\ve}{\left[\rho\left(t\right)\right.}\\&\left.
	-\rho\left(t_{0}\right)\right].}
	The simplified Eq. \eqref{eq:specialexpansion} holds for a number of systems, but we have yet to find an \textit{a priori} condition that guarantees its validity. Necessary and sufficient conditions for the vanishing of terms containing higher derivatives of the trace norm merit further study.
	
	\section{Negativity growth in various systems \label{sec:systems}}
	The derivatives of negativity can now be calculated by knowledge of the density matrix and its derivatives at a specific instant in time $t_0$. This is often much simpler than computing $\rho^{T_B}\left(t\right)$ for all times, finding all of the eigenvalues, and then differentiating the sum in Eq. \eqref{eq:negativity-eigvals}. Moreover, the latter method can be difficult to implement numerically, as it relies on derivatives of absolute value functions, which can lead to spurious results if not treated carefully. \color{black}Here we introduce some physical systems to exemplify the usefulness and robustness of our method.
	
	\subsection{Jaynes-Cummings model}
	A commonly used model in quantum optics is the Jaynes-Cummings model (JCM), which characterizes a two-level atom interacting with a single quantized mode of a bosonic field. The JCM has been the subject of much theoretical and experimental work \cite{ShoreKnight1993,Finketal2008}, including recent theoretical studies of its entanglement properties \cite{Quesada2013,Cresswell2017a}. The JCM Hamiltonian is given in units of $\hbar=1$ by
	\begin{equation}
		{H_\text{JCM}}{=}{\omega\id}\otimes{\had\ha}{ +}\left({\omega}{-}{\Delta}\right)\hat{\sigma}^\dagger\hat{\sigma}\otimes\id{-}\iu g\left(\hat{\sigma}\otimes\had{-}\hat{\sigma}^\dagger\otimes\ha\right),
	\end{equation}
	where $\ha$ is the bosonic annihilation operator for the field, $\hat{\sigma}=\ket{g}\bra{e}$ lowers the atom from the excited state $\ket{e}$ to the ground state $\ket{g}$, $\omega$ is the frequency of the bosonic mode, $\Delta$ is the detuning between the mode and the atomic transition frequency, and $g$ is a coupling constant
	\cite{Quesada2013}. The Hamiltonian conserves total excitation number $\id\otimes\had\ha+\hat{\sigma}^\dagger\hat{\sigma}\otimes\id$, restricting the dynamics to systems of size $2\times N$, where $N$ is the number of Fock states of the bosonic mode that are coupled to by the initial conditions. This subsumes systems for which the PPT criterion is sufficient ($N=2,\,3$), as well as systems that can have vanishing negativity yet remain entangled ($N>3$). We use this model to explore both types of systems, as delineated by the \hyperref[thm:PHC]{Peres-Horodecki Criterion}.
	
\begin{figure}\centering\includegraphics[trim={0cm 1.3cm 0cm 3.8cm
		},clip,width=1\columnwidth]{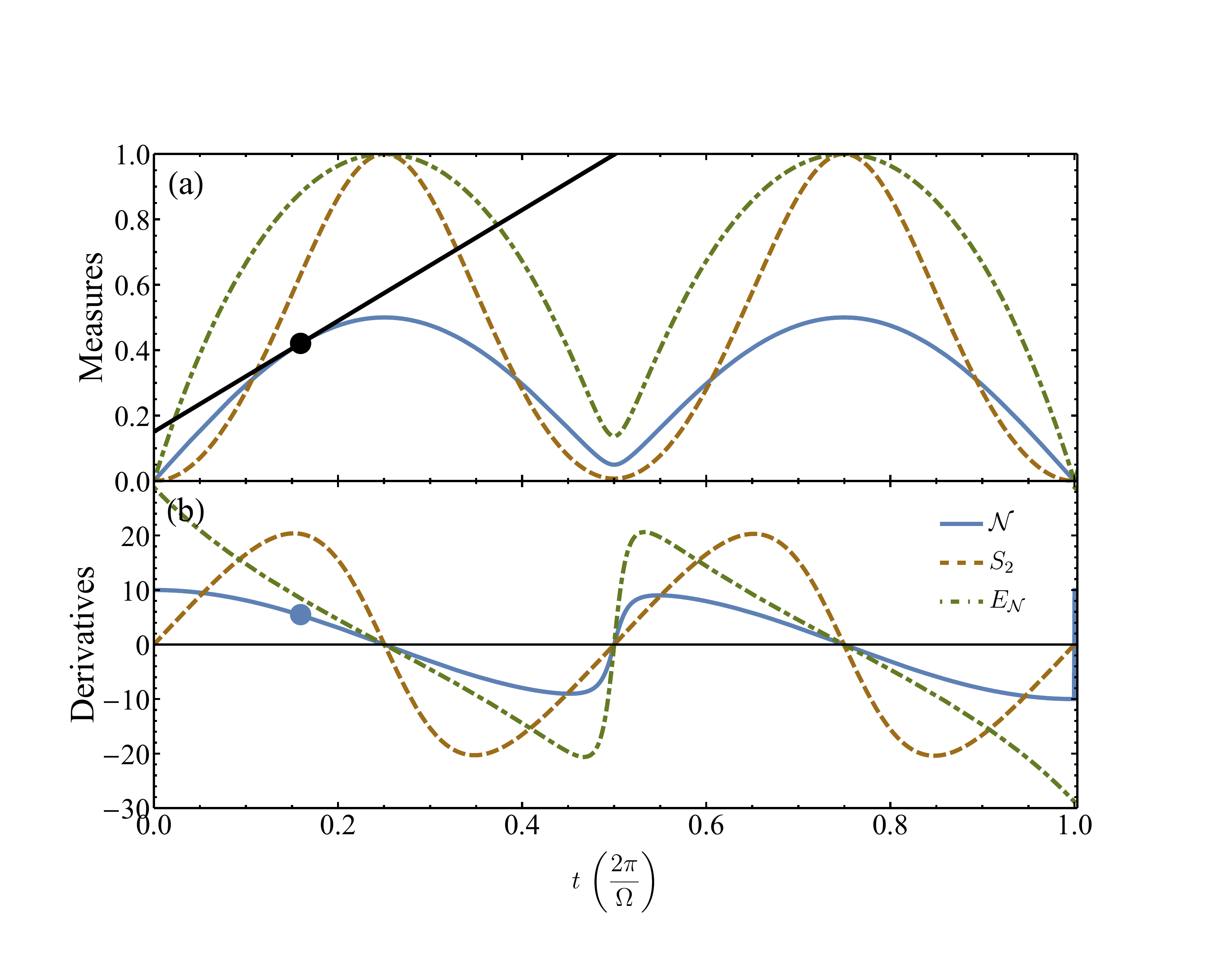}
		\caption{\small Evolution of the entanglement measures negativity $\mathcal{N}$ (blue solid lines), second-order R\'enyi entropy $S_2$ (brown dashed lines), and logarithmic negativity $E_{\mathcal{N}}$ (green dot-dashed lines) in the Jaynes-Cummings model for the initial state $\ket{e}\otimes\ket{3}$ and parameters $\left(\omega,\Delta,g\right)=\left(10,1,5\right)$. \textbf{(a)} Entanglement measures and \textbf{(b)} time-derivatives of entanglement measures versus time, with time in units of $2\pi/\Omega$. The time derivatives involving negativity are calculated using Eq. \eqref{eq:first derivative}. The measures agree in regions where entanglement is increasing, decreasing, maximal, minimal, and absent. Notably, negativity is {initially} more sensitive than R\'enyi entropies to growth in entanglement from the initial, separable state; the R\'enyi entropy does not grow linearly in time around $t=0$. We exemplify the success of  Eq. \eqref{eq:first derivative} by plotting a tangent to the negativity curve in (a) with slope found from the derivative curve in (b).\normalsize}
		\label{fig:JCM-2by2-measures}
	\end{figure}

	As a first example we choose the initial conditions ${\rho}(t_0)=\ket{\psi_0}\bra{\psi_0},\,\ket{\psi_0}=\ket{e}\otimes\ket{3}$; the atom is in its excited state and the field has three excitations. The state of the system for all time is given by $\rho(t)=\ket{\psi\left(t\right)}\bra{\psi\left(t\right)}$, where 
	\begin{align}
	    \begin{aligned}
	    {\ket{\psi\left(t\right)}}{=}&\eu^{-\iu H_\text{JCM}t}\ket{\psi_0}
		=\frac{\eu^{\iu t\left(\Delta-8\omega\right)/2}}{\Omega}\left[4g\sin\left(\frac{\Omega t}{2}\right)\ket{g}
		\otimes\ket{4}\right.\\&\left.
		+\left(\Omega\cos\frac{\Omega t}{2}-\iu \Delta\sin\frac{\Omega t}{2}\right)\ket{e}\otimes\ket{3}\right]
		,
	    \end{aligned}
	\end{align}
	and we have defined the Rabi frequency through $\Omega^2\coloneqq\Delta^2+\left(4g\right)^2$. The negativity can be calculated analytically for this system, which has an effective dimension of $2\times2$ at all times; the atom Hilbert space is spanned by $\ket{g}$ and $\ket{e}$ while the field Hilbert space is spanned by $\ket{3}$ and $\ket{4}$, with excitations trading between the two subsystems.
	We find, using \eqref{eq:first derivative}, that \eq{\frac{d\mathcal{N}}{d t}
		&=\frac{2g\sin\left(\Omega t\right)\left[\Delta^2+\left(4g\right)^2\cos\Omega t\right]}{\Omega\sqrt{\Delta^2\left(1-\cos\Omega t\right)^2+\Omega^2\sin^2\Omega t}},} which agrees with the result obtained by differentiating \eqref{eq:negativity-eigvals} with respect to time.
	We plot the negativity, second-order R\'enyi entropy, and logarithmic negativity for this system versus time for some fiducial parameters in Fig. \ref{fig:JCM-2by2-measures}. All of these quantities act as entanglement measures for the $2\times2$ system, as guaranteed by the PPT criterion. Of note, the measures involving negativity are initially more sensitive than the R\'enyi entropy, as the former grow linearly with time from separable states while the latter grows only quadratically \cite{Cresswell2017a}.

\begin{figure}\centering\includegraphics[width=\columnwidth]{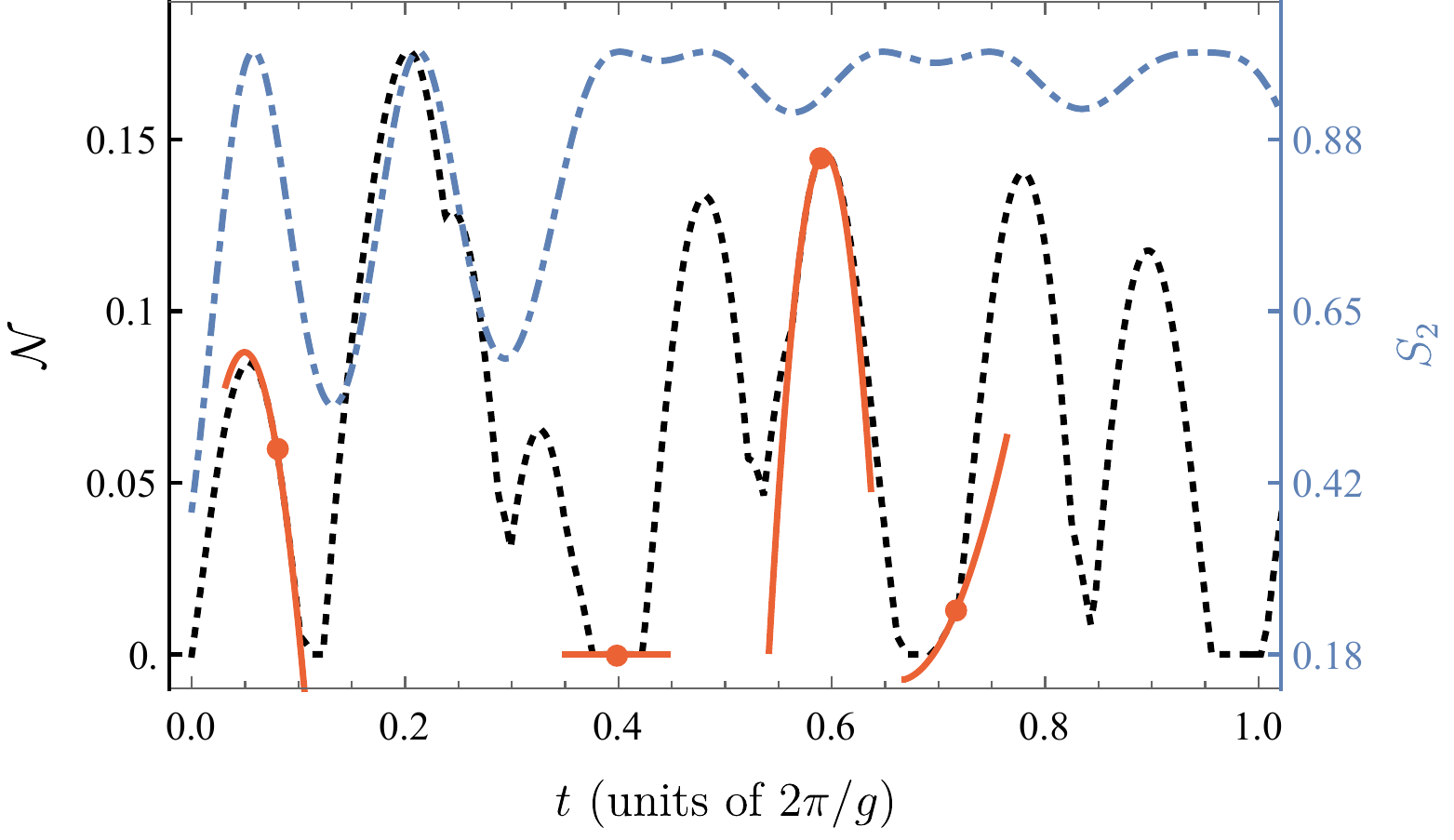}
		\caption{\small Evolution of entanglement monotones in the Jaynes-Cummings model for the initially bound-entangled state given by Eq. \eqref{eq:JCM-bound-initial} and parameters $\left(\omega,\Delta,g\right)=\left(10,1,5\right)$. The negativity (dotted black curve) and R\'enyi entropy (dot-dashed blue curve) no longer oscillate with a single frequency, so we measure time in units of $2\pi/g$. We plot the second-order expansion from Eqs. \eqref{eq:first derivative} and \eqref{eq:second derivative} about various time points (solid orange parabolas); these agree with numerically-calculated derivatives of Eq. \eqref{eq:negativity-eigvals} in regions where the graph of negativity is both concave and convex as a function of time. Moreover, at times when the R\'enyi entropy is changing yet negativity is constant, our method successfully captures the dynamics of negativity.
			\normalsize}	
		\label{fig:negativity-JCM-bound-renyi}
	\end{figure}

	To investigate a system for which the PPT criterion is not sufficient , let us choose the initial state
	\eq{
		\label{eq:JCM-bound-initial}
		\nonumber  \rho\propto &
		\ket{g}\bra{g}\otimes\left[4\left(\ket{0}\bra{0}+\ket{1}\bra{1}\right)+9\left(\ket{2}\bra{2}+\ket{3}\bra{3}\right)\right]\\
		&+\ket{e}\bra{e}\otimes\left(\ket{0}\bra{0}+\ket{1}\bra{1}+\ket{2}\bra{2}+\ket{3}\bra{3}\right)
		\\
		&+\left\lbrace\ket{g}\bra{e}\otimes\left[2\left(\ket{1}\bra{0}+\ket{2}\bra{1}\right)+3\ket{3}\bra{2}\right]+\text{H.c.}
		\right\rbrace,
	}
	which was shown in Ref. \cite{Quesada2013} to have zero negativity while remaining entangled (these states are `bound' entangled \cite{Horodeckietal1998Bound,Lavoieetal2010}).
	The negativity and second-order R\'enyi entropy are plotted in Fig. \ref{fig:negativity-JCM-bound-renyi} for the same fiducial parameters as in Fig. \ref{fig:JCM-2by2-measures}. There are distinct regions in which the negativity fails to witness entanglement, i.e., in which negativity is zero and R\'enyi entropy is nonzero (such as $t=0$). The first and second derivatives, given by Eqs. \eqref{eq:first derivative} and Eq. \eqref{eq:second derivative}, agree to machine precision with the results obtained by differentiating \eqref{eq:negativity-eigvals}. We also plot in Fig. \ref{fig:negativity-JCM-bound-renyi} the second-order expansion found using Eqs. \eqref{eq:first derivative} and \eqref{eq:second derivative} about an assortment of time points to show that our equations capture the negativity dynamics even in regions where negativity is constant,  \textcolor{black}{in intervals when $\rho^{T_B}$ is positive semi-definite,} and in the presence of bound entanglement. One may also use our method to analyze how negativity changes with respect to the system parameters $\Delta$ and $g$ in order to explore how entanglement in the JCM is sensitive to the entire parameter landscape.
	
	\subsection{Open system dynamics: entangled cavity photons}
	\label{sec:open system}
	
	The perturbation theory developed above admits variations of negativity  with respect to any parameter $\mu$, given the derivatives $\partial^n\rho/\partial\mu^n$. In the JCM examples, we used time as the perturbation parameter, with the unitary evolution equation $\partial\rho/\partial t=-\iu\left[H_\text{JCM},\rho\right]$. A natural extension of our method is to parametrize non-unitary evolution; we can ask how negativity changes with time in systems whose dynamics are coupled to other, external systems. Sometimes the external systems themselves are responsible for the entanglement generated with time \cite{BenattiFloreaniniPiani2003}. We can also ask how negativity changes with respect to other parameters, including dynamical parameters and initial conditions. In this section we exhibit the versatility of our method in another quantum-optical context.
	
\begin{figure}\centering\includegraphics[trim={0cm 2.0cm 0cm 3.4cm
		},clip,width=\columnwidth]{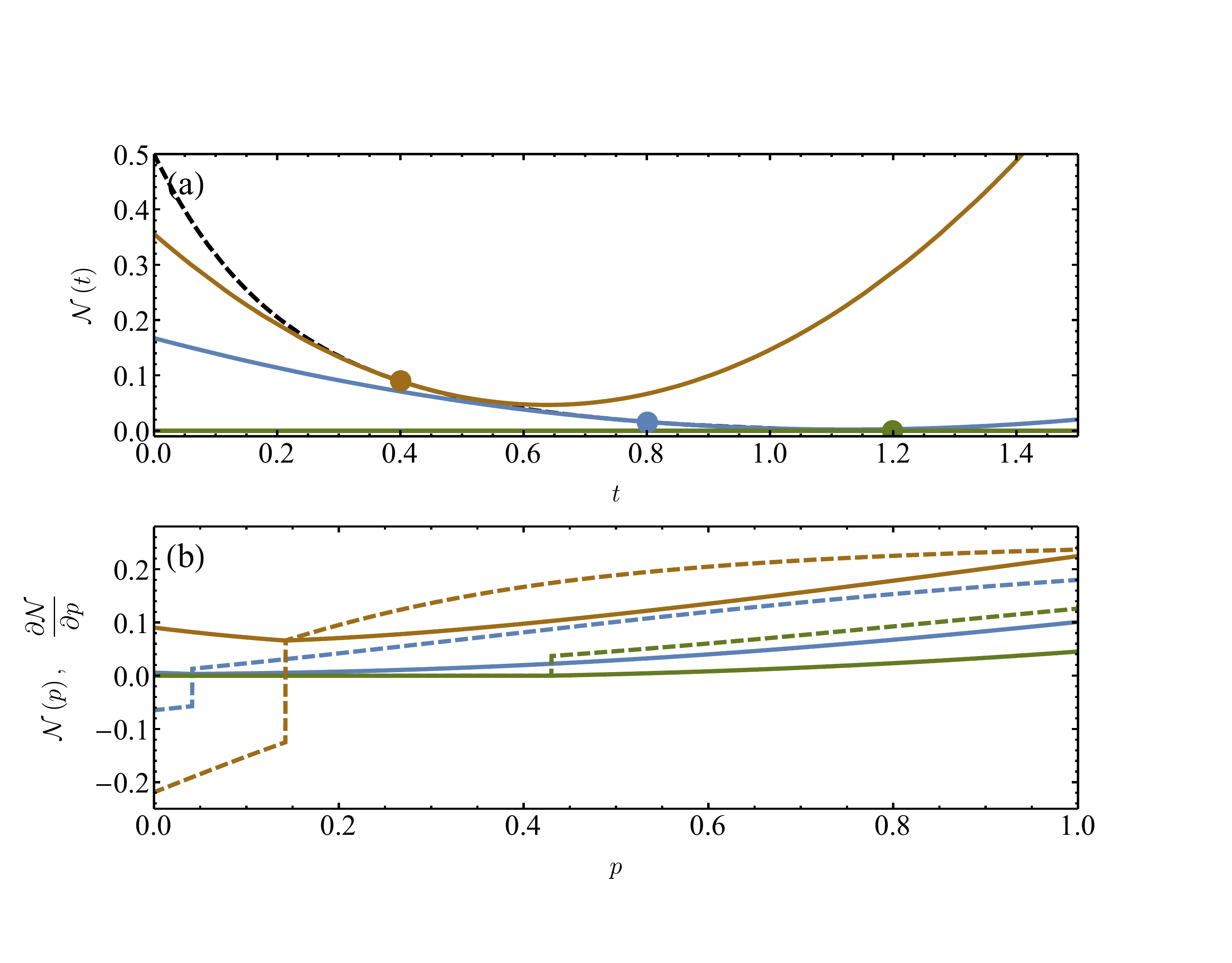}
		\caption{\small
			Dependence of negativity on system parameters in an open quantum system. \textbf{(a)} Evolution of negativity with respect to time (black, dashed curve), and perturbative expansions to second order using Eqs. \eqref{eq:first derivative} and \eqref{eq:second derivative} at time points $t=0.4$ (brown), $t=0.8$ (blue), and $t= 1.2$ (green). The initial mixing parameter is set to $p=0.35$. Negativity decays to 0 in finite time, beyond which all of the derivatives vanish, as successfully captured by our perturbation theory. Time is measured in units of the decay constant defined in Ref. \cite{Wangetal2016}. \textbf{(b)}
			Negativity (solid lines) and its derivatives with respect to the initial mixing parameter $p$ (dashed lines), for the same three time points as in (a) ($t=0.4$ is kinked at $p\approx 0.14$, $t=0.8$ at $p\approx 0.04$, and $t=1.2$ at $p\approx 0.43$). Regardless of time, negativity reaches a minimum for a particular value of $p$. When $t$ is small, negativity reaches a minimum for some $p=p_0$ between $0$ and $1$; $p_0$ is not monotonic with $t$. When $t$ is sufficiently large, negativity vanishes for all $p$ below a critical value, as seen in the $t=1.2$ curves. The derivatives are again calculated using Eq. \eqref{eq:first derivative}, and agree numerically with those found by differentiating $\mathcal{N}\left(p\right)$.\normalsize
		}
		\label{fig:arxiv}
	\end{figure}

	Negativity has recently been studied in the open system of a pair of cavities coupled to a pair of reservoirs with a flat spectrum {\cite{Lopezetal2008,Wangetal2016}}. The authors of Ref. \cite{Wangetal2016} showed that an initial mixture of maximally-entangled pairs of cavities, with states given by 
	\eq{\label{eq:cavity state}\rho_\text{cav}(t=0;&\,p)=p\ket{\psi}\bra{\psi}+\left(1-p\right)\ket{\phi}\bra{\phi},\quad 0\leq p\leq 1\\ &\ket{\psi}\propto\ket{0}\otimes\ket{0}+\ket{1}\otimes\ket{1},\\ &\ket{\phi}\propto\ket{0}\otimes\ket{2}+\ket{1}\otimes\ket{3},}
	can exhibit \emph{entanglement sudden death}; viz., negativity can decay to zero in finite time \cite{Almeidaetal2007,AlQasimiJames2008}. Furthermore, the cavity states coupled to by the dynamics are of dimension $2\times4$, so the PPT criterion does not hold in this system. The authors supply an analytic expression for $\rho_\text{cav}\left(t;\,p\right)$ (see App. \ref{app:open system dynamics} below), which can be compared to our perturbation theory method (Fig. \ref{fig:arxiv}). 
	
	In Fig. \ref{fig:arxiv}(a) we see a perfect agreement between our perturbation theory and the evolution of negativity with respect to time in this open system for a particular value of $p$. The dynamics are fully captured, including the time beyond which negativity decays to 0 and remains unchanged.
	
	Fig. \ref{fig:arxiv}(b) shows negativity and its derivatives from Eq. \eqref{eq:first derivative} with respect to the initial mixing parameter $p$ at various time points. The derivatives again match those found by differentiating Eq. \eqref{eq:negativity-eigvals} to machine precision. They give insight into the entanglement sudden death phenomenon, showing its dependence on initial conditions, as studied in depth in Ref. \cite{Wangetal2016}. Depending on the amount of initial mixing between the two entangled states $\ket{\psi_1}$ and $\ket{\psi_2}$, negativity decays at different rates with respect to time. For time evolution, negativity eventually reaches zero and remains there. With respect to $p$, negativity exhibits another sudden death feature: it decays to zero with shrinking $p$ at sufficiently long times. However, at shorter times, negativity reaches a minimum at intermediate values $p=p_0$, then grows again for increasing $\left|p-p_0\right|$, where the $p_0$ values are are highly sensitive to the time at which they are being evaluated. Our perturbation theory is an excellent tool for probing these complex phenomena or the dependence of negativity on any parameter $\mu$ in all systems for which $\partial\rho/\partial\mu$ is known.

	\section{Discussion \label{sec:Discussion}}

	In the preceding sections, we have provided a means of computing the
	perturbative expansion of the entanglement negativity. The complete
	expressions require knowledge of the partial transpose $\rho^{T_{B}}$
	at the expansion point; of the partial commutation matrix $K_B$, which we have presented explicitly in various convenient forms; and of the dynamics of $\rho$.
	
	Because the trace norm is not differentiable at points where $\rho^{T_{B}}$ is singular, we assumed that $\rho^{T_{B}}$ was invertible in our discussion. It would be interesting to know the conditions for $\rho^{T_B}$ to be invertible based on properties of $\rho$. There is, to our knowledge, no straightforward relationship between the rank of a general density matrix and the rank of its partial transpose. On the other hand, we can make some conclusions by considering pure states. Let $\sigma=\ket{\psi}\bra{\psi}$ be a pure state in a bipartite Hilbert space with dimensions $d_{A}\times d_{B}$ and let $r$ be the \emph{Schmidt rank} of $\ket{\psi}$, the number of non-zero coefficients in its Schmidt decomposition. From \cite{Johnston2018}, we know that the matrix rank of $\sigma^{T_{B}}$ is $r^{2}$. Since $r$ is bounded from above by the minimum of $\left\{ d_{A},d_{B}\right\}$, $\sigma^{T_{B}}$ having maximal rank implies that $r=d_{A}=d_{B}$. For pure states, then, $\sigma^{T_{B}} = \left(\ket{\psi}\bra{\psi}\right)^{T_{B}} $ is invertible if and only if the Schmidt rank of $\ket{\psi}$ is maximal and the dimensions of the subsystems are equal.
	
	For singular $\rho^{T_{B}}$  it might be possible to apply our analysis to the evolution of $\rho$ in an $r^{2}$- dimensional subspace where $\rho^{T_{B}}$ is supported. Otherwise, the trace norm still has a well defined subdifferential because it is convex \cite{Watson1992}. It may be possible to optimize over the set of subgradients given extra input, such as the global bound $\mathcal{N}\geq 0$, to determine the evolution of $\mathcal{N}(t)$ around singular points and find a one-sided derivative. We leave these ideas for future exploration.

	The primary challenge in our perturbative expansion was in the correct application of patterned matrix calculus to the problem. The salient pattern was the Hermiticity of the density operator, which implies Hermiticity of its partial transpose. We were able to extend the approach taken in \cite{Hjorungnes2011} to compute the first and second derivatives of the trace norm with respect to a Hermitian argument. However, Hermiticity may not be the only pattern at play. Density matrices are also normalized to have unit trace, and evolution may conspire to endow additional structure to the partial transpose. Furthermore, many calculations have been presented in the literature involving functions of density matrices that seemingly did not require patterned matrix calculus.
	
	As we have discussed, patterned derivatives can be found by first computing unpatterned derivatives, and subsequently imposing patterns on the result. Hermiticity is a strong condition which, as we show in App. \ref{sec:Acomp_pat}, destroys the independence of the complex differentials $d\ve \rho^{T_B}$ and $d\ve \left(\rho^{T_B}\right)^*$. Hence, imposing Hermiticity greatly alters the functional form of the derivatives, resulting in \eqref{eq:DTrNorm}.
	By constrast, the unit trace condition for $\rho^{T_{B}}$ introduces some dependencies among the diagonal elements, but this structure does not affect the patterned derivatives. The unit trace condition is a numerical constraint that does not change the functional form of the derivatives, and can simply be applied to the unpatterned derivative. For this reason we have not endeavoured to treat it with the same rigour as Hermiticity.

	Since we allow for general dynamics of $\rho(t)$, in theory its evolution might induce patterns on the partial transpose beyond Hermiticity. For example, one can conceive of a Hamiltonian that keeps $\rho^{T_{B}}$ positive semidefinite for some time interval, indicating a protracted separability or bound entanglement. In such a scenario the patterned derivatives may be functionally different from the Jacobian \eqref{eq:DTrNorm} or Hessian \eqref{eq:hessscal}, and would need to be treated on a case-by-case basis. However, we have seen in Section \ref{sec:open system} that our expansion correctly predicts zero evolution of the negativity when $\rho^{T_B}$ is positive semidefinite.

	More intriguing is to understand why the subtleties of patterned matrix calculus can be ignored in many calculations, yet are unavoidable when studying negativity. In App. \ref{sec:conjindcase} we show that, for complex analytic matrix functions, that is, for functions $G(X)$ that do not depend explicitly on the complex conjugate $X^*$, there is no functional difference between patterned and unpatterned derivatives. This coincidence allows one to gloss over the patterns of the density matrix when studying common functions like R\'enyi entropies $S_\alpha(\rho)=\tfrac{1}{1-\alpha}\log \tr \left(\rho^\alpha\right)$ \cite{Kim1996, Cresswell2017a}. In contrast, the trace norm $\left\Vert X\right\Vert_{1}=\tr \sqrt{X^\dag X}$ explicitly depends on the complex conjugate $X^*$, so that the negativity is not analytic in this sense. Some readers may also have been perturbed to notice that no consideration was given to the patterns of $\rho$ when computing $\partial\,\ve \rho^{T_{B}}/\partial\,\veT \rho$ in \eqref{eq:drhoTBdrho}. One way to explain this is that the function that maps $\ve \rho$ to its partial transpose $\ve \rho^{T_{B}}=K_B\ve \rho$ is constant, depending only on the dimensions of $\rho$, and hence is functionally independent of $\rho^{*}$.
	
	Our analysis can be applied to probe changes in negativity in a broad assortment of physical systems{, and the techniques we employ can be readily adapted to other functions of quantum states and observables}. Studies of phenomena as disparate as phase transitions \cite{Lu2018}, quantum quenches \cite{Coser2014,EislerZimboras2014}, and beam propagation \cite{Fedorovetal2018} can harness our methods in their investigations of negativity. 
	{One especially interesting application is to the linear, or nearly linear, growth of entanglement observed in a large class of many-body systems using entanglement entropy \cite{Calabrese2005,Bianchi2018} and, more recently, negativity \cite{Calabreseetal2012,Coser2014}. In critical systems, quasi-particles produced by a quench spread at a uniform velocity, leading to an emergent lightcone-like behaviour and exactly linear growth of logarithmic negativity. For more general systems, quasi-particles can propagate at varying speeds, leading to an approximate linear growth that has been studied numerically \cite{Coser2014,EislerZimboras2014}. Our approach to the derivatives of negativity provides a new avenue to analytically explore the conditions under which second and higher derivatives of the negativity will vanish. }
	
	{The techniques we employ can also be readily adapted to other functions of quantum states and observables.}
	For example, our perturbation theory can be applied to any dynamics involving the trace norm. This includes the trace distance $\tfrac{1}{2}\left\Vert \rho-\sigma\right\Vert_{1}$ between two states, which has been used, for instance, to investigate non-Markovian systems \cite{Breueretal2009,Rivasetal2010,Aaronsonetal2013}. {It has been shown that non-Markovianity holds when the trace distance between two states undergoing the same dynamics increases over time \cite{Breueretal2009}, a condition that can now be investigated using our matrix calculus techniques for non-analytic matrix functions. Fidelity, Hilbert-Schmidt distance, and other norm-based functions of the density matrix are all non-analytic in the sense discussed above, and can similarly be elucidated. One intriguing example is the application of our methods to quantum speed limits which usually involve bounds on $\left\Vert \dot \rho\right\Vert$ for some norm. However, some interesting speed limits can instead be expressed in terms of time derivative of $\left\Vert \rho\right\Vert$ \cite{Deffner2017}. This remains the topic of ongoing research for which our mathematical methods are well suited.}
	
	{Our techniques can even be employed for classical applications of complex patterned matrices, such as analyzing the condition number for Mueller matrices \cite{Layden2012}, whose patterns are discussed in \cite{Simon2010}}.
	Understanding the evolution of entanglement {and other functions of complex patterned matrices} will have ramifications for an expansive range of fields in the near future.

	\begin{acknowledgments}
		We would like to thank Eli Bourassa, Aharon Brodutch, Daniel James, Hoi-Kwong Lo, Hudson Pimenta, Erik Tonni, and Guifre Vidal for helpful discussions. JCC is supported by a Vanier Canada Graduate Scholarship and a Discovery Grant from the Natural Sciences and Engineering Research Council of Canada (NSERC). IT is supported by an Ontario Graduate Scholarship and an NSERC Discovery Grant. AZG is supported by the Alexander Graham Bell Scholarship \#504825 and NSERC Discovery Award Fund \#480483.
	\end{acknowledgments}

	\appendix
	
	\section{Vectorized representation of the partial transposition map\label{sec:superoperator}}
	
	\numberwithin{equation}{section}
	\setcounter{equation}{0}
	
	\color{black}Here we derive Eq. \eqref{eq:supop}, where the partial transposition
	map $T_B=\id\otimes T$ is recast to act on vectorized $d_{A}d_{B}\times d_{A}d_{B}$ matrices as in \eqref{eq:vectorizedsupop}, and takes the form of \eqref{eq153}, namely $K_B=\sum_{i}B^T_{i}\otimes A_{i}$. This is accomplished by finding the action of $T_B$ on each element of the standard basis of matrices, and then vectorizing. 
	
	The standard basis consists of single-entry matrices $(J^{ij})_{kl}=\delta_{ik}\delta_{jl}$ with the following ordering (we reserve
	$J^{ij}$ with no subscript for the $d_{A}d_{B}\times d_{A}d_{B}$ case):
	\begin{equation}
	\left\{J^{1,1} ,J^{2,1},\cdots,J^{d_{A}d_{B}-1,d_{A}d_{B}} ,J^{d_{A}d_{B},d_{A}d_{B}}\right\}.
	\end{equation}
	If we parametrize $i$ and $j$ by
	\begin{equation}
	i=(a_{i}-1)d_{B}+b_{i},\quad j=(a_{j}-1)d_{B}+b_{j},
	\end{equation}
	with $1\leq a_{i},a_{j}\leq d_{A}$ and $1\leq b_{i},b_{j}\leq d_{B}$, then we can decompose $J^{ij}$ as
	\begin{equation}
	J^{ij}=J^{(a_{i}-1)d_{B}+b_{i},(a_{j}-1)d_{B}+b_{j}}=J_{d_{A}}^{a_{i},a_{j}}\otimes J_{d_{B}}^{b_{i},b_{j}},
	\end{equation}
	using $m\times m$ single-entry matrices $J_{m}^{ij}$.
	From this we can read off the action of the partial transposition transformation on the basis elements
	\begin{equation}\label{eq:TBonMatrices}
	T_{B}\left(J^{ij}\right)=J_{d_{A}}^{a_{i},a_{j}}\otimes J_{d_{B}}^{b_{j},b_{i}}=J^{(a_{i}-1)d_{B}+b_{j},(a_{j}-1)d_{B}+b_{i}}.
	\end{equation}
	
	In the vectorized representation we use the basis
	\begin{equation}
	\{\ve J^{1,1} ,\ve J^{2,1},{\cdots},\ve J^{d_{A}d_{B}-1,d_{A}d_{B}} ,\ve J^{d_{A}d_{B},d_{A}d_{B}}\},
	\end{equation}
	and the action of $T_B$ on basis elements is determined by vectorizing both sides of \eqref{eq:TBonMatrices} such that $ \ve T_{B}\left(J^{ij}\right) = K_B \ve J^{ij}$.
	$K_B$ is a $\left(d_Ad_B\right)^2\times \left(d_Ad_B\right)^2$ permutation matrix whose elements can be expressed in terms of the single-entry matrices $J^{qr}_{\left(d_Ad_B\right)^2}$, with $1\leq q,r\leq \left(d_Ad_B\right)^2$. Note that these matrices are larger than the matrices $J^{ij}$ with no subscripts. The $\ve J^{ij}$ basis element has its non-zero entry in position $r=(j-1)d_{A}d_{B}+i$. Hence, the $r^{\text{th}}$ column of $K_B$ is equal to $\ve T_B\left(J^{ij}\right)$. From \eqref{eq:TBonMatrices} we see that partial transposition takes $i\to i^\prime=\left(a_i-1\right)d_B+b_j$ and $j\to j^\prime=\left(a_j-1\right)d_B+b_i$, so the $r^{\text{th}}$ column only has a non-zero entry in the $q^\text{th}$ row, where $q=\left(j^\prime-1\right)d_{A}d_{B}+i^\prime$. 
	This non-zero element of $K_B$ can be expressed as
	\begin{widetext}
	\begin{align}
	\begin{aligned}
	I(i,j)=
	&
	J_{\left(d_{A}d_{B}\right)^{2}}^{(j^\prime-1)d_{A}d_{B}+i^\prime,(j-1)d_{A}d_{B}+i}  = 
	J^{j^\prime,j}\otimes J^{i^\prime,i}
	=J^{(a_{j}-1)d_{B}+b_{i},(a_{j}-1)d_{B}+b_{j}}\otimes J^{(a_{i}-1)d_{B}+b_{j},(a_{i}-1)d_{B}+b_{i}}.
	\end{aligned}
	\end{align}
	
	Every vectorized basis element $\ve J^{ij}$ matches with an element $I\left(i,j\right)$, so that $K_B$ is the sum of all such elements:
	\begin{align}
	\begin{aligned}
	K_B= & \sum_{a_{i},a_{j}=1}^{d_{A}}\sum_{b_{i},b_{j}=1}^{d_{B}}J^{(a_{j}-1)d_{B}+b_{i},(a_{j}-1)d_{B}+b_{j}}\otimes J^{(a_{i}-1)d_{B}+b_{j},(a_{i}-1)d_{B}+b_{i}}\\
	= & \sum_{b_{i},b_{j}=1}^{d_{B}}\left(\sum_{a=1}^{d_{A}}J^{(a-1)d_{B}+b_{i},(a-1)d_{B}+b_{j}}\right)\otimes\left(\sum_{a=1}^{d_{A}}J^{(a-1)d_{B}+b_{j},(a-1)d_{B}+b_{i}}\right)\\
	= & \sum_{b_{i},b_{j}=1}^{d_{B}}\left(\sum_{a=1}^{d_{A}}J_{d_{A}}^{a,a}\otimes J_{d_{B}}^{b_{i},b_{j}}\right)\otimes\left(\sum_{a=1}^{d_{A}}J_{d_{A}}^{a,a}\otimes J_{d_{B}}^{b_{j},b_{i}}\right)\\
	= & \sum_{b_{i},b_{j}=1}^{d_{B}}\left(\id_{d_{A}}\otimes J_{d_{B}}^{b_{i},b_{j}}\right)\otimes\left(\id_{d_{A}}\otimes J_{d_{B}}^{b_{j},b_{i}}\right).
	\end{aligned}
	\end{align}
	\end{widetext}
	Another representation of $K_B$ involves a more optimal basis choice. Consider a symmetric matrix, $E$, an
	antisymmetric matrix, $O$, and an arbitrary matrix, $X$. Notice that
	\begin{align}
	T_{B}\left(X\otimes E\right) & =X\otimes E^{T}=X\otimes E,\\
	T_{B}\left(X\otimes O\right) & =X\otimes O^{T}=-X\otimes O.
	\end{align}
	Therefore the partial transposition map has as its eigenvectors matrices
	of the form $X\otimes E$ (eigenvalue 1) and $X\otimes O$ (eigenvalue
	-1). With this in mind, we can define bases $\mathbb{E}_{\text{S}}$ and
	$\mathbb{E}_{\text{A}}$ of the symmetric and antisymmetric matrices, respectively:
	\begin{align}
	\mathbb{E}_{\text{S}} & =\left\{ J^{ij}_{d_B}+J^{ji}_{d_B}-2J^{ij}_{d_B}J^{ij}_{d_B}\right\}, \\
	\mathbb{E}_{\text{AS}} & =\left\{ J^{ij}_{d_B}-J^{ji}_{d_B}\vert i \neq j \right\},
	\end{align}
	which gives us a basis for the combined system
	\begin{equation}
	\mathbb{E}=\left\{ J^{ij}_{d_A}\right\} \otimes\left(\mathbb{E}_{\text{S}}\cup\mathbb{E}_{\text{AS}}\right).
	\end{equation}
	$K_B$ is diagonal in the (vectorized) basis $\mathbb{E}$; if we assume the basis matrices are all normalized by their Frobenius norm and order the basis so that the symmetric matrices come first, it has the form
	\begin{equation}
	K_B^{\mathbb{E}}=\begin{bmatrix}\id_{k} & 0\\
	0 & -\id_{l}
	\end{bmatrix},
	\end{equation}
	where $\id_{k}$ is the $k \times k$ identity matrix with $k = d^{2}_{A}\left|\mathbb{E}_{\text{S}}\right|=\frac{1}{2}d^{2}_{A}d_{B}\left(d_{B}+1\right)$, and $\id_{l}$ is the $l \times l$ identity matrix with
	$l = d^{2}_{A}\left|\mathbb{E}_{\text{AS}}\right|=\frac{1}{2}d^{2}_{A}d_{B}\left(d_{B}-1\right)$.
	We can thus write
	\begin{equation}
	K_B=V^{\mathbb{E}}K_B^{\mathbb{E}}\left(V^{\mathbb{E}}\right)^{T},
	\end{equation}
	where $V^{\mathbb{E}}$ has as its column vectors the vectorized matrices
	from $\mathcal{\mathbb{E}}$.

	\section{Calculus of complex patterned matrices}
	
	In this appendix we recount the calculus of complex, patterned matrices developed in \cite{ Hjorungnes2007,Tracy1988,Hjorungnes2008,Hjorungnes2008a} and summarized in \cite{Hjorungnes2011}. {This calculus was used in \cite{Hjorungnes2011} to rigorously compute the derivative of a scalar function with respect to a Hermitian argument \eqref{eq:DTrNorm}, and also provides a rigourous derivation of our result on patterned Hessians \eqref{eq:hessscal}. We use these techniques in App. \ref{sec:conjindcase} to discuss the types of functions for which patterned and unpatterned derivatives are not equal.}
	
	\subsection{Formulation \label{sec:Acomp_pat}}
	
	Consider $F\left(P,W,W^{*}\right)$,
	a differentiable, complex matrix-valued function of a real matrix variable $P$,
	a complex matrix variable \textbf{$W$} and its complex conjugate
	$W^{*}$. The differential of such a function is given by
	\begin{align}\label{eq:differential}
	\begin{aligned}
	d\ve F=&\left(D_{P}F\right)d\ve P+\left(D_{W}F\right)d\ve W\\&+\left(D_{W^{*}}F\right)d\ve W^{*},
	\end{aligned}
	\end{align}
	where the differentials of $P$, $W$ and $W^{*}$ are independent,
	and the Jacobian $D_{P}F$, for example, is
	\begin{align}
	D_{P}F & \coloneqq\frac{\partial\,\ve F}{\partial\,\veT P}.
	\end{align}
	Fortuitously, the differential commutes with vectorization, tracing,
	transposition, and conjugation:
	\begin{align}
	\begin{aligned}
	d\left(\ve X\right)&=\ve\left(dX\right),\quad d\left(\tr X\right)=\tr dX,\\
	d\left(X^{T}\right)&=\left(dX\right)^{T},\quad d\left(X^{*}\right)=\left(dX\right)^{*}.
	\end{aligned}
	\end{align}
	Also, derivatives in this formalism satisfy a chain rule; for a composite function
	\begin{equation}\label{eq:chainrule}
	H\left(P,W,W^{*}\right)=G\left[F\left(P,W,W^{*}\right),F^{*}\left(P,W,W^{*}\right)\right],
	\end{equation}
	we have
	\begin{align}
	D_{P}H&=\left(D_{F}G\right)\left(D_{P}F\right)+\left(D_{F^{*}}G\right)\left(D_{P}F^{*}\right),\\
	D_{W}H&=\left(D_{F}G\right)\left(D_{W}F\right)+\left(D_{F^{*}}G\right)\left(D_{W}F^{*}\right),\\
	D_{W^*}H&=\left(D_{F}G\right)\left(D_{W^*}F\right)+\left(D_{F^{*}}G\right)\left(D_{W^*}F^{*}\right).
	\end{align}

	We must employ a careful strategy for taking derivatives with respect to a matrix if there are any elements in that matrix which are (possibly constant) functions of the other elements. Such an approach
	was developed in \cite{Tracy1988, Hjorungnes2008, Hjorungnes2008a} and we summarize it here for a differentiable function $G(A,A^*)$ of a complex patterned matrix $A$:
	
		(1) Let $F$ be a function that acts on a set of unpatterned matrices $[P,W,W^{*}]$
		to make a patterned matrix $A=F\left(P,W,W^{*}\right)$. This function
		must be differentiable with respect to $P,$ $W$, and $W^{*}$,
		and a diffeomorphism between the sets of patterned and unpatterned
		matrices, i.e., a smooth, bijective function whose inverse is also
		smooth{, must exist}. The number of independent parameters contained in $P,W,W^{*}$ that fully parametrize
		the set of patterned matrices should be minimal.
	
		(2) Let $X$ be an unpatterned matrix with the same size as $A$. Extend $G$ to act on unpatterned matrices and find its derivatives $D_{X}G\left(X,X^{*}\right)$ and $D_{X^*}G\left(X,X^{*}\right)$. Use the chain rule for $G(A,A^*)=H\left(P,W,W^{*}\right)$ as in \eqref{eq:chainrule} to find
		\begin{equation}
		{D_{P}H} {=}{D_{X}G}{\left({X}{,}{X^{*}}\right)}{\vert_{{X}{=}{A}}}D_{P}F{+}D_{X^{*}}G{\left({X}{,}{X^{*}}\right)}{\vert_{{X}{=}{A}}}D_{P}F^{*}, \label{eq:patapp}
		\end{equation}
		etc., where patterns are applied after differentiation.

	(3) The derivative of $G(A,A^*)$ with respect to the patterned matrix $A$ is given by
		\begin{equation}
		D_{A}G=\left[D_{P}H,D_{W}H,D_{W^{*}}H\right]D_{A}F^{-1}.\label{eq:finalpatapp}
		\end{equation}

	In other words, one should find a minimal basis to represent the set of patterned matrices $A$, compute derivatives in this basis, then transform back to the standard basis. The diffeomorphism $F(P,W,W^*)$ represents the transformation from the minimal basis to the standard basis, while its inverse $F^{-1}(A)$ produces a vector of matrices $[P,W,W^{*}]^T$.

	\color{black}

	\subsection{Analytic functions of matrices} \label{sec:conjindcase}
	In this section we discuss a sufficient condition for when consideration of matrix patterns is unnecessary in taking derivatives of functions with respect to those matrices. Let $G(X,X^{*})$ be a matrix-valued function that is differentiable in matrices $X$. Suppose that $G(X,X^{*})$ is analytic in the sense that it is independent of $X^{*}$, i.e., $D_{X^{*}}G=0$. Let $A$ be a patterned matrix with the same size as $X$, and $F$ an appropriate diffeomorphism acting on the minimal set of matrix parameters $\left[P,W,W^{*}\right]$ such that $A=F(P,W,W^{*})$. Define $H$ so that $H(P,W,W^{*})=G(A,A^{*})$ as in the beginning of App. \ref{sec:Acomp_pat}. We have, from Eq. \eqref{eq:patapp}, that
	\begin{equation}
	D_{P}H  =D_{X}G\left(X,X^{*}\right)\vert_{X=A}D_{P}F,
	\end{equation}
	and similarly for $D_{W}H$ and $D_{W^{*}}H$. This means that Eq. \eqref{eq:finalpatapp} now reads
	\begin{equation}
	D_{A}G = D_{X}G \vert _{X=A} [D_{P}F,D_{W}F,D_{W^{*}}F]D_{A}F^{-1}.
	\end{equation}
	But by construction the diffeomorphism $F$ satisfies,
	\begin{equation}
	[D_{P}F,D_{W}F,D_{W^{*}}F]D_{A}F^{-1}=\id,
	\end{equation}
	since this amounts to changing from the standard basis to the minimal basis and back \cite{Hjorungnes2011}.
	Thus we can see that
	\begin{equation}
	D_{A}G = D_{X}G \vert _{X=A}.
	\end{equation}
	
	The conclusion we draw is as follows: for functions independent of the complex conjugate of their argument, taking the patterned derivative is equivalent to differentiating with respect to the unpatterned argument and evaluating it at the patterned matrix. The trace norm does not obey this condition, and we found its patterned derivative to have a different form compared to the unpatterned counterparts.

	\section{Simplifying the Hessian of the trace norm}
	\label{sec:simphess}
	To obtain simplified expressions for the trace norm Hessian, we use identity \eqref{Kcommid} as well as a commutation rule for the matrix $K$ and the inverse of a Kronecker sum. Supposing $X \oplus Y$ is invertible, and remembering that $K$ is self-inverse, we have
	\begin{widetext}
	\begin{equation}
	K(X \oplus Y)^{-1} = [(X \oplus Y)K]^{-1} =[K(Y \oplus X)]^{-1} = (Y \oplus X)^{-1}K.
	\end{equation}
	We can now write down the $B$ matrices introduced in Eq. \eqref{eq:Aexpansion} using \eqref{eq:dvec|X|} in \eqref{eq:d2tr},
	\begin{align}
	B_{00}=&\left(\left|X\right|^{-1}\right)^{T}\otimes\id-\left(\id\otimes X\right)\left(\left|X\right|^{-1}\right)_{\oplus}\left(\id\otimes\left|X\right|^{-1}\right)\left(\left|X\right|^{-1}\right)_{\oplus}\left(\id\otimes X^{\dagger}\right),\\
	B_{01}=&-K\left(X\otimes\id\right)\left(\left|X\right|^{-1}\right)_{\oplus}^{T}\left(\left|X\right|^{-1}\otimes\id\right)\left(\left|X\right|^{-1}\right)_{\oplus}^{T}\left(\id\otimes X^{T}\right),\\
	B_{10}=&-K\left(X^{*}\otimes\id\right)\left(\left|X\right|^{-1}\right)_{\oplus}\left(\id\otimes\left|X\right|^{-1}\right)\left(\left|X\right|^{-1}\right)_{\oplus}\left(\id\otimes X^{\dagger}\right),\\
	B_{11}=&-\left(\id\otimes X^{*}\right)\left(\left|X\right|^{-1}\right)_{\oplus}^{T}\left(\left|X\right|^{-1}\otimes\id\right)\left(\left|X\right|^{-1}\right)_{\oplus}^{T}\left(\id\otimes X^{T}\right).
	\end{align}
	The unpatterned Hessians are then
	\begin{align}
	&\hess_{X,X}\left(\left\Vert X\right\Vert _{1}\right)=-\frac{1}{2}K\left(X^{*}\otimes\id\right)\left(|X|_{\oplus}\right)^{-1}\left(\left|X\right|^{-1}\right)_{\oplus}\left(|X|_{\oplus}\right)^{-1}\left(\id\otimes X^{\dagger}\right),\\
	&\hess_{X^{*},X^{*}}\left(\left\Vert X\right\Vert _{1}\right)=-\frac{1}{2}K\left(X\otimes\id\right)\left(|X|_{\oplus}^{T}\right)^{-1}\left(\left|X\right|^{-1}\right)_{\oplus}^{T}\left(|X|_{\oplus}^{T}\right)^{-1}\left(\id\otimes X^{T}\right),\\
	\hess_{X,X^{*}}\left(\left\Vert X\right\Vert _{1}\right)=&\frac{1}{2}\left(\left|X\right|^{-1}\right)^{T}\otimes\id-\frac{1}{2}\left(\id\otimes X\right)\left(|X|_{\oplus}\right)^{-1}\left(\left|X\right|^{-1}\right)_{\oplus}\left(|X|_{\oplus}\right)^{-1}\left(\id\otimes X^{\dagger}\right)=\left(\hess_{X^{*},X}\left\Vert X\right\Vert _{1}\right)^{T}.
	\end{align}
	These are combined to form the patterned Hessian with respect to a Hermitian matrix $A$ \eqref{eq:hessscal},
	\begin{equation}
	\hess_{A,A}\left(\left\Vert A\right\Vert _{1}\right) = \frac{1}{2}K\left[\left(\left|X\right|^{-1}\right)_{\oplus}-X_{\oplus}\left(\left|X\right|_{\oplus}\right)^{-1}\left(\left|X\right|^{-1}\right)_{\oplus}\left(\left|X\right|_{\oplus}\right)^{-1}X_{\oplus}^{\dagger}\right]_{X=A},
	\end{equation}
	which was presented in \eqref{eq:patternedHessian}. We can simplify this equation with the eigendecomposition $A=U\Lambda U^{\dagger}$ by noting
	\begin{align}
	A_\oplus&=A^{T}\oplus A=\left(U^{*}\otimes U\right)\left(\Lambda\oplus\Lambda\right)\left(U^{T}\otimes U^{\dagger}\right)\\
	\left\vert A\right\vert_\oplus&=\left|A\right|^{T}\oplus\left|A\right|=\left(U^{*}\otimes U\right)\left(\left|\Lambda\right|\oplus\left|\Lambda\right|\right)\left(U^{T}\otimes U^{\dagger}\right),
	\end{align}
	and so on. Continuing in this manner, the patterned Hessian can be written solely in terms of the eigendecomposition as
	\begin{align}\label{eq:altPatternedHessian}
	\hess_{A,A}\left(\left\Vert A\right\Vert _{1}\right)=&
	\frac{1}{2}K(U^*\otimes U)\left[
	(\vert \Lambda\vert^{-1}\oplus \vert \Lambda\vert^{-1})
	-(\Lambda\oplus \Lambda)\left(\vert \Lambda\vert\oplus\vert \Lambda\vert\right)^{-1}(\vert \Lambda\vert^{-1}\oplus\vert \Lambda\vert^{-1})\left(\vert \Lambda\vert\oplus\vert \Lambda\vert\right)^{-1}(\Lambda\oplus \Lambda)\right](U^T\otimes U^\dag)\\
	=&\frac{1}{2}K(U^*\otimes U)\left[
	\left(\vert \Lambda\vert\oplus\vert \Lambda\vert\right)^{2}
	-(\Lambda\oplus \Lambda)^2\right]\left(\vert \Lambda\vert\oplus\vert \Lambda\vert\right)^{-2}(\vert \Lambda\vert^{-1}\oplus \vert \Lambda\vert^{-1})(U^T\otimes U^\dag)\\
	=&K(U^*\otimes U)\left[
	\left|\Lambda\right|\otimes\left|\Lambda\right|-\Lambda\otimes\Lambda\right]\left(\vert \Lambda\vert\oplus\vert \Lambda\vert\right)^{-2}(\vert \Lambda\vert^{-1}\oplus \vert \Lambda\vert^{-1})(U^T\otimes U^\dag)\\
	=&K(U^*\otimes U)\left[
	\id-\text{sign}\Lambda\otimes\text{sign}\Lambda\right]\left(\left|\Lambda\right|\otimes\left|\Lambda\right|\right)\left(\vert \Lambda\vert\oplus\vert \Lambda\vert\right)^{-2}(\vert \Lambda\vert^{-1}\oplus \vert \Lambda\vert^{-1})(U^T\otimes U^\dag)\\
	=&K(U^*\otimes U)\left[
	\id-\text{sign}\Lambda\otimes\text{sign}\Lambda\right]\left(\vert \Lambda\vert\oplus\vert \Lambda\vert\right)^{-1}(U^T\otimes U^\dag),
	\end{align} where we relied on the fact that all of the matrices involving $\Lambda$ are diagonal and commute. The final line was presented in \eqref{eq:spectHessian2}.
	\color{black}
	
	\section{Details of open system dynamics}
	\label{app:open system dynamics}
	As per Ref. \cite{Wangetal2016}, the initial state of the pair of cavities is given by Eq. \eqref{eq:cavity state}, and each cavity is coupled to a reservoir with $N\to\infty$ modes. Defining two amplitudes $\xi\left(t\right)=\eu^{-t/2}$ and $\chi\left(t\right)=\sqrt{1-\eu^{-t}}$, where $t$ is measured in units of some dissipative constant, the state evolves to
	\eq{\rho\left(t;\,p\right)=\left(
		\begin{array}{cccccccc}
			a_{11} & 0 & 0 & 0 & 0 & a_{16} & 0 & 0 \\
			0 & a_{22} & 0 & 0 & 0 & 0 & a_{27} & 0 \\
			0 & 0 & a_{33} & 0 & 0 & 0 & 0 & a_{38} \\
			0 & 0 & 0 & a_{44} & 0 & 0 & 0 & 0 \\
			0 & 0 & 0 & 0 & a_{55} & 0 & 0 & 0 \\
			a_{16} & 0 & 0 & 0 & 0 & a_{66} & 0 & 0 \\
			0 & a_{27} & 0 & 0 & 0 & 0 & a_{77} & 0 \\
			0 & 0 & a_{38} & 0 & 0 & 0 & 0 & a_{88} \\
		\end{array}
		\right)
	}
	in the $\left\{\ket{0}\otimes\ket{0},\ket{0}\otimes\ket{1},\ket{0}\otimes\ket{2},\ket{0}\otimes\ket{3},\ket{1}\otimes\ket{0},\ket{1}\otimes\ket{1},\ket{1}\otimes\ket{2},\ket{1}\otimes\ket{3}\right\}$ basis, where the matrix elements are given by
	\begin{align}
	\begin{aligned}
		a_{11}=&\left(p+\chi^4+\chi^8-\chi^8\right)/2, \
		a_{22}=\xi^2\chi^2\left[2-p+3\left(1-p\right)\chi^4\right]/2, \
		a_{33}=\left(1-p\right)\xi^4\left(1+3\chi^4\right)/2, \\
		a_{44}=&\left(1-p\right)\xi^6\chi^2 / 2, \
		a_{55}=\xi^2\chi^2\left(p+\chi^4-p\chi^4\right)/2, \
		a_{66}=\xi^4\left[p+3\left(1-p\right)\chi^4\right]/2, \
		a_{77}=3\left(1-p\right)\xi^6\chi^2 / 2, \\
		a_{88}=&\left(1-p\right)\xi^8/2, \
		a_{16}=\xi^2\left[p+\sqrt{3}\left(1-p\right)\chi^4\right]/2, \
		a_{27}=\sqrt{3/2}\left(1-p\right)\xi^4\chi^2, \
		a_{38}=\left(1-p\right)\xi^6/2.
	\end{aligned}
	\end{align}
This can be used to calculate $\partial\rho/\partial t$ and $\partial\rho/\partial p$ in Eqs. \eqref{eq:first derivative} and \eqref{eq:second derivative}, and can also be used to explicitly calculate the eigenvalues of $\rho$ for use in Eq. \eqref{eq:negativity-eigvals}

	\end{widetext}
	\providecommand{\href}[2]{#2}\begingroup\raggedright\endgroup


\begin{thebibliography}{10}
		
		\bibitem{Horodecki2009}
		R.~Horodecki, P.~Horodecki, M.~Horodecki, and K.~Horodecki, ``Quantum
		entanglement,'' \href{http://dx.doi.org/10.1103/RevModPhys.81.865}{{\em Rev.
				Mod. Phys.} {\bfseries 81} (Jun, 2009) 865--942},
		{arXiv}:\href{http://arxiv.org/abs/quant-ph/0702225}{{\ttfamily
				quant-ph/0702225}}.
		
		\bibitem{Bennett1993}
		C.~H. Bennett, G.~Brassard, C.~Cr\'epeau, R.~Jozsa, A.~Peres, and W.~K.
		Wootters, ``Teleporting an unknown quantum state via dual classical and
		{E}instein-{P}odolsky-{R}osen channels,''
		\href{http://dx.doi.org/10.1103/PhysRevLett.70.1895}{{\em Phys. Rev. Lett.}
			{\bfseries 70} (Mar, 1993) 1895--1899}.
		
		\bibitem{Bouwmeester1997}
		D.~Bouwmeester, J.-W. Pan, K.~Mattle, M.~Eibl, H.~Weinfurter, and A.~Zeilinger,
		``Experimental quantum teleportation,''  {\em Nature} {\bfseries 390} (Dec.,
		1997) 575.
		
		\bibitem{Barrett2004}
		M.~D. Barrett, J.~Chiaverini, T.~Schaetz, J.~Britton, W.~M. Itano, J.~D. Jost,
		E.~Knill, C.~Langer, D.~Leibfried, R.~Ozeri, and D.~J. Wineland,
		``Deterministic quantum teleportation of atomic qubits,''  {\em Nature}
		{\bfseries 429} (June, 2004) 737.
		
		\bibitem{Riebe2004}
		M.~Riebe, H.~H\"affner, C.~F. Roos, W.~H\"ansel, J.~Benhelm, G.~P.~T.
		Lancaster, T.~W. K\:orber, C.~Becher, F.~Schmidt-Kaler, D.~F.~V. James, and
		R.~Blatt, ``Deterministic quantum teleportation with atoms,''  {\em Nature}
		{\bfseries 429} (June, 2004) 734.
		
		\bibitem{Mitchell2004}
		M.~W. Mitchell, J.~S. Lundeen, and A.~M. Steinberg, ``Super-resolving phase
		measurements with a multiphoton entangled state,'' {\em Nature} {\bfseries
			429} (May, 2004) 161,
		{arXiv}:\href{http://arxiv.org/abs/quant-ph/0312186}{{\ttfamily
				quant-ph/0312186}}.
		
		\bibitem{Pezze2009}
		L.~Pezz\'e and A.~Smerzi, ``Entanglement, nonlinear dynamics, and the
		{H}eisenberg limit,''
		\href{http://dx.doi.org/10.1103/PhysRevLett.102.100401}{{\em Phys. Rev.
				Lett.} {\bfseries 102} (Mar, 2009) 100401},
		{arXiv}:\href{http://arxiv.org/abs/0711.4840}{{\ttfamily 0711.4840}}.
		
		\bibitem{giovannetti2011advances}
		V.~Giovannetti, S.~Lloyd, and L.~Maccone, ``Advances in quantum metrology,''
		{\em Nature Photonics} {\bfseries 5} (Mar., 2011) 222,
		{arXiv}:\href{http://arxiv.org/abs/1102.2318}{{\ttfamily 1102.2318}}.
		
		\bibitem{gross2012nonlinear}
		C.~Gross, T.~Zibold, E.~Nicklas, J.~Est\`eve, and M.~K. Oberthaler, ``Nonlinear
		atom interferometer surpasses classical precision limit,'' {\em Nature}
		{\bfseries 464} (Mar., 2010) 1165,
		{arXiv}:\href{http://arxiv.org/abs/1009.2374}{{\ttfamily 1009.2374}}.
		
		\bibitem{Ekert1991}
		A.~K. Ekert, ``Quantum cryptography based on {B}ell's theorem,''
		\href{http://dx.doi.org/10.1103/PhysRevLett.67.661}{{\em Phys. Rev. Lett.}
			{\bfseries 67} (Aug, 1991) 661--663}.
		
		\bibitem{Gisin2002}
		N.~Gisin, G.~Ribordy, W.~Tittel, and H.~Zbinden, ``Quantum cryptography,''
		\href{http://dx.doi.org/10.1103/RevModPhys.74.145}{{\em Rev. Mod. Phys.}
			{\bfseries 74} (Mar, 2002) 145--195},
		{arXiv}:\href{http://arxiv.org/abs/quant-ph/0101098}{{\ttfamily
				quant-ph/0101098}}.
		
		\bibitem{Curty2004}
		M.~Curty, M.~Lewenstein, and N.~L\"utkenhaus, ``Entanglement as a precondition
		for secure quantum key distribution,''
		\href{http://dx.doi.org/10.1103/PhysRevLett.92.217903}{{\em Phys. Rev. Lett.}
			{\bfseries 92} (May, 2004) 217903},
		{arXiv}:\href{http://arxiv.org/abs/quant-ph/0307151}{{\ttfamily
				quant-ph/0307151}}.
		
		\bibitem{Tang2014}
		Z.~Tang, Z.~Liao, F.~Xu, B.~Qi, L.~Qian, and H.-K. Lo, ``Experimental
		demonstration of polarization encoding measurement-device-independent quantum
		key distribution,''
		\href{http://dx.doi.org/10.1103/PhysRevLett.112.190503}{{\em Phys. Rev.
				Lett.} {\bfseries 112} (May, 2014) 190503},
		{arXiv}:\href{http://arxiv.org/abs/1306.6134}{{\ttfamily 1306.6134}}.
		
		\bibitem{Bennett1996}
		C.~H. Bennett, D.~P. DiVincenzo, J.~A. Smolin, and W.~K. Wootters,
		``Mixed-state entanglement and quantum error correction,''
		\href{http://dx.doi.org/10.1103/PhysRevA.54.3824}{{\em Phys. Rev. A}
			{\bfseries 54} (Nov, 1996) 3824--3851},
		{arXiv}:\href{http://arxiv.org/abs/quant-ph/9604024}{{\ttfamily
				quant-ph/9604024}}.
		
		\bibitem{Wootters1998}
		W.~K. Wootters, ``Entanglement of formation of an arbitrary state of two
		qubits,'' \href{http://dx.doi.org/10.1103/PhysRevLett.80.2245}{{\em Phys.
				Rev. Lett.} {\bfseries 80} (Mar, 1998) 2245--2248},
		{arXiv}:\href{http://arxiv.org/abs/quant-ph/9709029}{{\ttfamily
				quant-ph/9709029}}.
		
		\bibitem{Renyi1961}
		A.~R{\'e}nyi, ``On measures of information and entropy,''  {\em Proceedings of
			the fourth Berkeley Symposium on Mathematics, Statistics and Probability
			1960} (1961) 547--561.
		
		\bibitem{Plenio2005}
		M.~B. Plenio and S.~Virmani, ``An introduction to entanglement measures,'' {\em
			Quantum Information \& Computation} {\bfseries 7} no.~1, (2007) 1--51,
		{arXiv}:\href{http://arxiv.org/abs/quant-ph/0504163v3}{{\ttfamily
				quant-ph/0504163v3}}.
		
		\bibitem{Peres1996}
		A.~Peres, ``Separability criterion for density matrices,''
		\href{http://dx.doi.org/10.1103/PhysRevLett.77.1413}{{\em Phys. Rev. Lett.}
			{\bfseries 77} (Aug, 1996) 1413--1415},
		{arXiv}:\href{http://arxiv.org/abs/quant-ph/9604005}{{\ttfamily
				quant-ph/9604005}}.
		
		\bibitem{Vollbrecht2002}
		K.~G.~H. Vollbrecht and M.~M. Wolf, ``Conditional entropies and their relation
		to entanglement criteria,'' \href{http://dx.doi.org/10.1063/1.1498490}{{\em
				Journal of Mathematical Physics} {\bfseries 43} no.~9, (2002) 4299--4306},
		{arXiv}:\href{http://arxiv.org/abs/quant-ph/0202058}{{\ttfamily
				quant-ph/0202058}}.
		
		\bibitem{Horodecki1996}
		M.~Horodecki, P.~Horodecki, and R.~Horodecki, ``Separability of mixed states:
		necessary and sufficient conditions,''
		\href{http://dx.doi.org/https://doi.org/10.1016/S0375-9601(96)00706-2}{{\em
				Physics Letters A} {\bfseries 223} no.~1, (1996) 1 -- 8},
		{arXiv}:\href{http://arxiv.org/abs/quant-ph/9605038}{{\ttfamily
				quant-ph/9605038}}.
		
		\bibitem{Zyczkowski1998}
		K.~\ifmmode~\dot{Z}\else \.{Z}\fi{}yczkowski, P.~Horodecki, A.~Sanpera, and
		M.~Lewenstein, ``Volume of the set of separable states,''
		\href{http://dx.doi.org/10.1103/PhysRevA.58.883}{{\em Phys. Rev. A}
			{\bfseries 58} (Aug, 1998) 883--892},
		{arXiv}:\href{http://arxiv.org/abs/quant-ph/9804024}{{\ttfamily
				quant-ph/9804024}}.
		
		\bibitem{Vidal2001}
		G.~Vidal and R.~F. Werner, ``Computable measure of entanglement,''
		\href{http://dx.doi.org/10.1103/PhysRevA.65.032314}{{\em Phys. Rev. A}
			{\bfseries 65} (Feb, 2002) 032314},
		{arXiv}:\href{http://arxiv.org/abs/quant-ph/0102117}{{\ttfamily
				quant-ph/0102117}}.
		
		\bibitem{Plenio2005a}
		M.~B. Plenio, ``Logarithmic negativity: A full entanglement monotone that is
		not convex,'' \href{http://dx.doi.org/10.1103/PhysRevLett.95.090503}{{\em
				Phys. Rev. Lett.} {\bfseries 95} (Aug, 2005) 090503},
		{arXiv}:\href{http://arxiv.org/abs/quant-ph/0505071}{{\ttfamily
				quant-ph/0505071}}.
		
		\bibitem{Bennett1996a}
		C.~H. Bennett, H.~J. Bernstein, S.~Popescu, and B.~Schumacher, ``Concentrating
		partial entanglement by local operations,''
		\href{http://dx.doi.org/10.1103/PhysRevA.53.2046}{{\em Phys. Rev. A}
			{\bfseries 53} (Apr, 1996) 2046--2052},
		{arXiv}:\href{http://arxiv.org/abs/quant-ph/9511030}{{\ttfamily
				quant-ph/9511030}}.
		
		\bibitem{Huang2014}
		Y.~Huang, ``Computing quantum discord is np-complete,'' {\em New Journal of
			Physics} {\bfseries 16} no.~3, (2014) 033027,
		{arXiv}:\href{http://arxiv.org/abs/1305.5941}{{\ttfamily 1305.5941}}.
		
		\bibitem{Audenaert2003}
		K.~Audenaert, M.~B. Plenio, and J.~Eisert, ``Entanglement cost under
		positive-partial-transpose-preserving operations,''
		\href{http://dx.doi.org/10.1103/PhysRevLett.90.027901}{{\em Phys. Rev. Lett.}
			{\bfseries 90} (Jan, 2003) 027901},
		{arXiv}:\href{http://arxiv.org/abs/quant-ph/0207146}{{\ttfamily
				quant-ph/0207146}}.
		
		\bibitem{Audenaertetal2002}
		K.~Audenaert, J.~Eisert, M.~B. Plenio, and R.~F. Werner, ``Entanglement
		properties of the harmonic chain,''
		\href{http://dx.doi.org/10.1103/PhysRevA.66.042327}{{\em Phys. Rev. A}
			{\bfseries 66} (Oct, 2002) 042327},
		{arXiv}:\href{http://arxiv.org/abs/quant-ph/0205025}{{\ttfamily
				quant-ph/0205025}}.
		
		\bibitem{Anders2008}
		J.~Anders, ``Thermal state entanglement in harmonic lattices,''
		\href{http://dx.doi.org/10.1103/PhysRevA.77.062102}{{\em Phys. Rev. A}
			{\bfseries 77} (Jun, 2008) 062102},
		{arXiv}:\href{http://arxiv.org/abs/0803.1102}{{\ttfamily 0803.1102}}.
		
		\bibitem{EislerZimboras2014}
		V.~Eisler and Z.~Zimbor�s, ``Entanglement negativity in the harmonic chain out
		of equilibrium,'' {\em New Journal of Physics} {\bfseries 16} no.~12, (2014)
		123020,  {arXiv}:\href{http://arxiv.org/abs/1406.5474}{{\ttfamily
				1406.5474}}.
		
		\bibitem{HelmesWessel2014}
		J.~Helmes and S.~Wessel, ``Entanglement entropy scaling in the bilayer
		{H}eisenberg spin system,''
		\href{http://dx.doi.org/10.1103/PhysRevB.89.245120}{{\em Phys. Rev. B}
			{\bfseries 89} (Jun, 2014) 245120},
		{arXiv}:\href{http://arxiv.org/abs/1403.7395}{{\ttfamily 1403.7395}}.
		
		\bibitem{EislerZimboras2015}
		V.~Eisler and Z.~Zimbor{\'a}s, ``On the partial transpose of fermionic
		{G}aussian states,'' {\em New Journal of Physics} {\bfseries 17} no.~5,
		(2015) 053048,  {arXiv}:\href{http://arxiv.org/abs/1502.01369}{{\ttfamily
				1502.01369}}.
		
		\bibitem{Shermanetal2016}
		N.~E. Sherman, T.~Devakul, M.~B. Hastings, and R.~R.~P. Singh,
		``Nonzero-temperature entanglement negativity of quantum spin models: {A}rea
		law, linked cluster expansions, and sudden death,''
		\href{http://dx.doi.org/10.1103/PhysRevE.93.022128}{{\em Phys. Rev. E}
			{\bfseries 93} (Feb, 2016) 022128},
		{arXiv}:\href{http://arxiv.org/abs/1510.08005}{{\ttfamily 1510.08005}}.
		
		\bibitem{Shimetal2018}
		J.~Shim, H.-S. Sim, and S.-S.~B. Lee, ``Numerical renormalization group method
		for entanglement negativity at finite temperature,''
		\href{http://dx.doi.org/10.1103/PhysRevB.97.155123}{{\em Phys. Rev. B}
			{\bfseries 97} (Apr, 2018) 155123},
		{arXiv}:\href{http://arxiv.org/abs/1808.08506}{{\ttfamily 1808.08506}}.
		
		\bibitem{Calabreseetal2012}
		P.~Calabrese, J.~Cardy, and E.~Tonni, ``Entanglement negativity in quantum
		field theory,'' \href{http://dx.doi.org/10.1103/PhysRevLett.109.130502}{{\em
				Phys. Rev. Lett.} {\bfseries 109} (Sep, 2012) 130502},
		{arXiv}:\href{http://arxiv.org/abs/1206.3092}{{\ttfamily 1206.3092}}.
		
		\bibitem{Coser2014}
		A.~Coser, E.~Tonni, and P.~Calabrese, ``Entanglement negativity after a global
		quantum quench,'' {\em Journal of Statistical Mechanics: Theory and
			Experiment} {\bfseries 2014} no.~12, (2014) P12017,
		{arXiv}:\href{http://arxiv.org/abs/1410.0900}{{\ttfamily 1410.0900}}.
		
		\bibitem{Chaturvedietal2018}
		P.~Chaturvedi, V.~Malvimat, and G.~Sengupta, ``Holographic quantum entanglement
		negativity,'' \href{http://dx.doi.org/10.1007/JHEP05(2018)172}{{\em Journal
				of High Energy Physics} {\bfseries 2018} no.~5, (May, 2018) 172},
		{arXiv}:\href{http://arxiv.org/abs/1609.06609}{{\ttfamily 1609.06609}}.
		
		\bibitem{Kim1996}
		J.~I. Kim, M.~C. Nemes, A.~F.~R. de~Toledo~Piza, and H.~E. Borges,
		``Perturbative expansion for coherence loss,''
		\href{http://dx.doi.org/10.1103/PhysRevLett.77.207}{{\em Phys. Rev. Lett.}
			{\bfseries 77} (Jul, 1996) 207--210}.
		
		\bibitem{Yang2017}
		I.-S. Yang, ``Entanglement timescale,''
		\href{http://dx.doi.org/10.1103/PhysRevD.97.066008}{{\em Phys. Rev. D}
			{\bfseries 97} (Mar, 2018) 066008},
		{arXiv}:\href{http://arxiv.org/abs/1707.05792}{{\ttfamily 1707.05792}}.
		
		\bibitem{Cresswell2017a}
		J.~C. Cresswell, ``Universal entanglement timescale for {R}\'{e}nyi
		entropies,'' \href{http://dx.doi.org/10.1103/PhysRevA.97.022317}{{\em Phys.
				Rev. A} {\bfseries 97} (Feb, 2018) 022317},
		{arXiv}:\href{http://arxiv.org/abs/1709.10064}{{\ttfamily 1709.10064}}.
		
		\bibitem{Watson1992}
		G.~Watson, ``Characterization of the subdifferential of some matrix norms,''
		\href{http://dx.doi.org/10.1016/0024-3795(92)90407-2}{{\em Linear Algebra and
				its Applications} {\bfseries 170} (1992) 33 -- 45}.
		
		\bibitem{Hjorungnes2011}
		A.~Hj{\o}rungnes, {\em Complex-Valued Matrix Derivatives With Applications in
			Signal Processing and Communications}.
		\newblock Cambridge University Press,
		\newblock 2011.
		
		\bibitem{Magnus1985}
		J.~R. Magnus and H.~Neudecker, ``Matrix differential calculus with applications
		to simple, {H}adamard, and {K}ronecker products,''
		\href{http://dx.doi.org/https://doi.org/10.1016/0022-2496(85)90006-9}{{\em
				Journal of Mathematical Psychology} {\bfseries 29} no.~4, (1985) 474 -- 492}.
		
		\bibitem{Magnus2007}
		J.~R. Magnus and H.~Neudecker, {\em Matrix Differential Calculus with
			Applications in Statistics and Econometrics}.
		\newblock Wiley, 3rd~ed.,
		\newblock 2007.
		
		\bibitem{Havel2002}
		T.~F. Havel, ``Robust procedures for converting among {L}indblad, {K}raus and
		matrix representations of quantum dynamical semigroups,''
		\href{http://dx.doi.org/10.1063/1.1518555}{{\em Journal of Mathematical
				Physics} {\bfseries 44} no.~2, (2003) 534--557},
		{arXiv}:\href{http://arxiv.org/abs/quant-ph/0201127}{{\ttfamily
				quant-ph/0201127}}.
		
		\bibitem{Zyczkowski2004}
		K.~{\.{Z}}yczkowski and I.~Bengtsson, ``On duality between quantum maps and
		quantum states,''
		\href{http://dx.doi.org/10.1023/B:OPSY.0000024753.05661.c2}{{\em Open Systems
				{\&} Information Dynamics} {\bfseries 11} no.~1, (Mar, 2004) 3--42},
		{arXiv}:\href{http://arxiv.org/abs/quant-ph/0401119}{{\ttfamily
				quant-ph/0401119}}.
		
		\bibitem{Jaroslaw2011}
		J.~A. Miszcak, ``Singular value decomposition and matrix reorderings in quantum
		information theory,'' \href{http://dx.doi.org/10.1142/S0129183111016683}{{\em
				International Journal of Modern Physics C} {\bfseries 22} no.~09, (2011)
			897--918},  {arXiv}:\href{http://arxiv.org/abs/1011.1585}{{\ttfamily
				1011.1585}}.
		
		\bibitem{Hjorungnes2007}
		A.~Hj{\o}rungnes and D.~Gesbert, ``Complex-valued matrix differentiation:
		Techniques and key results,''
		\href{http://dx.doi.org/10.1109/TSP.2007.893762}{{\em IEEE Transactions on
				Signal Processing} {\bfseries 55} no.~6, (June, 2007) 2740--2746}.
		
		\bibitem{Higham2008}
		N.~J. Higham, {\em Functions of Matrices: Theory and Computation}, vol.~104.
		\newblock SIAM,
		\newblock 2008.
		
		\bibitem{ShoreKnight1993}
		B.~W. Shore and P.~L. Knight, ``The {J}aynes-{C}ummings model,''
		\href{http://dx.doi.org/10.1080/09500349314551321}{{\em Journal of Modern
				Optics} {\bfseries 40} no.~7, (1993) 1195--1238}.
		
		\bibitem{Finketal2008}
		J.~M. Fink, M.~G{\"o}ppl, M.~Baur, R.~Bianchetti, P.~J. Leek, A.~Blais, and
		A.~Wallraff, ``Climbing the {J}aynes-{C}ummings ladder and observing its
		nonlinearity in a cavity {QED} system,'' {\em Nature} {\bfseries 454} (July,
		2008) 315,  {arXiv}:\href{http://arxiv.org/abs/0902.1827}{{\ttfamily
				0902.1827}}.
		
		\bibitem{Quesada2013}
		N.~Quesada and A.~Sanpera, ``Bound entanglement in the {J}aynes-{C}ummings
		model,'' {\em Journal of Physics B: Atomic, Molecular and Optical Physics}
		{\bfseries 46} no.~22, (2013) 224002,
		{arXiv}:\href{http://arxiv.org/abs/1305.2604}{{\ttfamily 1305.2604}}.
		
		\bibitem{Horodeckietal1998Bound}
		M.~Horodecki, P.~Horodecki, and R.~Horodecki, ``Mixed-state entanglement and
		distillation: Is there a ``bound'' entanglement in nature?,''
		\href{http://dx.doi.org/10.1103/PhysRevLett.80.5239}{{\em Phys. Rev. Lett.}
			{\bfseries 80} (Jun, 1998) 5239--5242},
		{arXiv}:\href{http://arxiv.org/abs/quant-ph/9801069}{{\ttfamily
				quant-ph/9801069}}.
		
		\bibitem{Lavoieetal2010}
		J.~Lavoie, R.~Kaltenbaek, M.~Piani, and K.~J. Resch, ``Experimental bound
		entanglement in a four-photon state,''
		\href{http://dx.doi.org/10.1103/PhysRevLett.105.130501}{{\em Phys. Rev.
				Lett.} {\bfseries 105} (Sep, 2010) 130501},
		{arXiv}:\href{http://arxiv.org/abs/1005.1258}{{\ttfamily 1005.1258}}.
		
		\bibitem{BenattiFloreaniniPiani2003}
		F.~Benatti, R.~Floreanini, and M.~Piani, ``Environment induced entanglement in
		markovian dissipative dynamics,''
		\href{http://dx.doi.org/10.1103/PhysRevLett.91.070402}{{\em Phys. Rev. Lett.}
			{\bfseries 91} (Aug, 2003) 070402},
		{arXiv}:\href{http://arxiv.org/abs/quant-ph/0307052}{{\ttfamily
				quant-ph/0307052}}.
		
		\bibitem{Lopezetal2008}
		C.~E. L\'opez, G.~Romero, F.~Lastra, E.~Solano, and J.~C. Retamal, ``Sudden
		birth versus sudden death of entanglement in multipartite systems,''
		\href{http://dx.doi.org/10.1103/PhysRevLett.101.080503}{{\em Phys. Rev.
				Lett.} {\bfseries 101} (Aug, 2008) 080503},
		{arXiv}:\href{http://arxiv.org/abs/0802.1825}{{\ttfamily 0802.1825}}.
		
		\bibitem{Wangetal2016}
		L.-D. Wang, L.-T. Wang, M.~Yang, J.-Z. Xu, Z.~D. Wang, and Y.-K. Bai,
		``Entanglement and measurement-induced nonlocality of mixed maximally
		entangled states in multipartite dynamics,''
		\href{http://dx.doi.org/10.1103/PhysRevA.93.062309}{{\em Phys. Rev. A}
			{\bfseries 93} (Jun, 2016) 062309},
		{arXiv}:\href{http://arxiv.org/abs/1506.06878}{{\ttfamily 1506.06878}}.
		
		\bibitem{Almeidaetal2007}
		M.~P. Almeida, F.~de~Melo, M.~Hor-Meyll, A.~Salles, S.~P. Walborn, P.~H.~S.
		Ribeiro, and L.~Davidovich, ``Environment-induced sudden death of
		entanglement,''  \href{http://dx.doi.org/10.1126/science.1139892}{{\em
				Science} {\bfseries 316} no.~5824, (2007) 579--582}.
		
		\bibitem{AlQasimiJames2008}
		A.~Al-Qasimi and D.~F.~V. James, ``Sudden death of entanglement at finite
		temperature,'' \href{http://dx.doi.org/10.1103/PhysRevA.77.012117}{{\em Phys.
				Rev. A} {\bfseries 77} (Jan, 2008) 012117},
		{arXiv}:\href{http://arxiv.org/abs/0707.2611}{{\ttfamily 0707.2611}}.
		
		\bibitem{Johnston2018}
		N.~Johnston and E.~Patterson, ``The inverse eigenvalue problem for entanglement
		witnesses,'' \href{http://dx.doi.org/10.1016/j.laa.2018.03.043}{{\em Linear
				Algebra and its Applications} {\bfseries 550} (2018) 1 -- 27},
		{arXiv}:\href{http://arxiv.org/abs/1708.05901}{{\ttfamily 1708.05901}}.
		
		\bibitem{Lu2018}
		T.-C. Lu and T.~Grover, ``Singularity in entanglement negativity across finite
		temperature phase transitions,''
		{arXiv}:\href{http://arxiv.org/abs/1808.04381}{{\ttfamily 1808.04381}}.
		
		\bibitem{Fedorovetal2018}
		K.~G. Fedorov, S.~Pogorzalek, U.~Las~Heras, M.~Sanz, P.~Yard, P.~Eder,
		M.~Fischer, J.~Goetz, E.~Xie, K.~Inomata, Y.~Nakamura, R.~Di~Candia,
		E.~Solano, A.~Marx, F.~Deppe, and R.~Gross, ``Finite-time quantum
		entanglement in propagating squeezed microwaves,'' {\em Scientific Reports}
		{\bfseries 8} no.~1, (Apr., 2018) 6416,
		{arXiv}:\href{http://arxiv.org/abs/1703.05138}{{\ttfamily 1703.05138}}.
		
		\bibitem{Calabrese2005}
		P.~Calabrese and J.~Cardy, ``Evolution of entanglement entropy in
		one-dimensional systems,'' {\em Journal of Statistical Mechanics: Theory and
			Experiment} {\bfseries 2005} no.~04, (2005) P04010,
		{arXiv}:\href{http://arxiv.org/abs/cond-mat/0503393}{{\ttfamily
				cond-mat/0503393}}.
		
		\bibitem{Bianchi2018}
		E.~Bianchi, L.~Hackl, and N.~Yokomizo, ``Linear growth of the entanglement
		entropy and the kolmogorov-sinai rate,''
		\href{http://dx.doi.org/10.1007/JHEP03(2018)025}{{\em Journal of High Energy
				Physics} {\bfseries 2018} no.~3, (Mar, 2018) 25},
		{arXiv}:\href{http://arxiv.org/abs/1709.00427}{{\ttfamily 1709.00427}}.
		
		\bibitem{Breueretal2009}
		H.-P. Breuer, E.-M. Laine, and J.~Piilo, ``Measure for the degree of
		non-markovian behavior of quantum processes in open systems,''
		\href{http://dx.doi.org/10.1103/PhysRevLett.103.210401}{{\em Phys. Rev.
				Lett.} {\bfseries 103} (Nov, 2009) 210401},
		{arXiv}:\href{http://arxiv.org/abs/0908.0238}{{\ttfamily 0908.0238}}.
		
		\bibitem{Rivasetal2010}
		A.~Rivas, S.~F. Huelga, and M.~B. Plenio, ``Entanglement and non-markovianity
		of quantum evolutions,''
		\href{http://dx.doi.org/10.1103/PhysRevLett.105.050403}{{\em Phys. Rev.
				Lett.} {\bfseries 105} (Jul, 2010) 050403},
		{arXiv}:\href{http://arxiv.org/abs/0911.4270}{{\ttfamily 0911.4270}}.
		
		\bibitem{Aaronsonetal2013}
		B.~Aaronson, R.~L. Franco, G.~Compagno, and G.~Adesso, ``Hierarchy and dynamics
		of trace distance correlations,'' {\em New Journal of Physics} {\bfseries 15}
		no.~9, (2013) 093022,
		{arXiv}:\href{http://arxiv.org/abs/1307.3953}{{\ttfamily 1307.3953}}.
		
		\bibitem{Deffner2017}
		S.~Deffner and S.~Campbell, ``Quantum speed limits: from heisenberg's
		uncertainty principle to optimal quantum control,'' {\em Journal of Physics
			A: Mathematical and Theoretical} {\bfseries 50} no.~45, (2017) 453001,
		{arXiv}:\href{http://arxiv.org/abs/1705.08023}{{\ttfamily 1705.08023}}.
		
		\bibitem{Layden2012}
		D.~Layden, M.~F.~G. Wood, and I.~A. Vitkin, ``Optimum selection of input
		polarization states in determining the sample mueller matrix: a dual
		photoelastic polarimeter approach,''
		\href{http://dx.doi.org/10.1364/OE.20.020466}{{\em Opt. Express} {\bfseries
				20} no.~18, (Aug, 2012) 20466--20481}.
		
		\bibitem{Simon2010}
		B.~N. Simon, S.~Simon, F.~Gori, M.~Santarsiero, R.~Borghi, N.~Mukunda, and
		R.~Simon, ``Nonquantum entanglement resolves a basic issue in polarization
		optics,'' \href{http://dx.doi.org/10.1103/PhysRevLett.104.023901}{{\em Phys.
				Rev. Lett.} {\bfseries 104} (Jan, 2010) 023901},
		{arXiv}:\href{http://arxiv.org/abs/0906.2467}{{\ttfamily 0906.2467}}.
		
		\bibitem{Tracy1988}
		D.~S. Tracy and K.~G. Jinadasa, ``Patterned matrix derivatives,''  {\em The
			Canadian Journal of Statistics / La Revue Canadienne de Statistique}
		{\bfseries 16} no.~4, (1988) 411--418.
		
		\bibitem{Hjorungnes2008}
		A.~Hj{\o}rungnes and D.~P. Palomar,
		\href{http://dx.doi.org/10.1109/SAM.2008.4606875}{``Patterned complex-valued
			matrix derivatives,''} in {\em 2008 5th IEEE Sensor Array and Multichannel
			Signal Processing Workshop}, pp.~293--297.
		\newblock
		\newblock July, 2008.
		
		\bibitem{Hjorungnes2008a}
		A.~Hj{\o}rungnes and D.~P. Palomar,
		\href{http://dx.doi.org/10.1109/ISABEL.2008.4712619}{``Finding patterned
			complex-valued matrix derivatives by using manifolds,''} in {\em 2008 First
			International Symposium on Applied Sciences on Biomedical and Communication
			Technologies}, pp.~1--5.
		\newblock
		\newblock Oct, 2008.
		
	\end{thebibliography}
\end{document}